\documentclass[10pt,journal,compsoc]{IEEEtran}
\ifCLASSOPTIONcompsoc
  \usepackage[nocompress]{cite}
\else
  \usepackage{cite}
\fi
\ifCLASSINFOpdf
  \usepackage[pdftex]{graphicx}
  \graphicspath{{./pdf/}{./jpeg/}{./png/}{./jpg/}}
  \DeclareGraphicsExtensions{.pdf,.jpeg,.png,.jpg}
\else
\fi

\ifCLASSOPTIONcaptionsoff
 \usepackage[nomarkers]{endfloat}
\let\MYoriglatexcaption\caption
\renewcommand{\caption}[2][\relax]{\MYoriglatexcaption[#2]{#2}}
\fi
\usepackage{silence}
\usepackage[T1]{fontenc}
\usepackage[utf8]{inputenc}
\usepackage[french,british]{babel}
\usepackage{ragged2e}

\usepackage{algorithm2e}
\usepackage{listings}
\usepackage{color}
\usepackage{xcolor}
\usepackage{fancyvrb}
\usepackage{url}
\usepackage{verbatim}
\usepackage{multirow}
\usepackage{pgfplots}
\usepackage{pgfplotstable}
\usepackage{rotating}
\usetikzlibrary{positioning}

\usepackage{amsmath}
\usepackage{booktabs}
\usepackage{threeparttable}
\usepackage{hyperref}

\usepackage{blindtext}
\usepackage{scrextend}
\addtokomafont{labelinglabel}{\sffamily}

\usepackage{adjustbox}
\newcommand{\ra}[1]{\renewcommand{\arraystretch}{#1}}

\definecolor{dkgreen}{rgb}{0,0.6,0}
\definecolor{gray}{rgb}{0.5,0.5,0.5}
\definecolor{mauve}{rgb}{0.58,0,0.82}
\usepackage{inconsolata}

\definecolor{pblue}{rgb}{0.13,0.13,1}
\definecolor{pgreen}{rgb}{0,0.5,0}
\definecolor{pred}{rgb}{0.9,0,0}
\definecolor{pgrey}{rgb}{0.46,0.45,0.48}
\definecolor{cverbbg}{gray}{0.93}

\lstdefinelanguage{mytemplate}{
  xleftmargin=0.8em,
  framexleftmargin= 0.5em,
  frame=single,
  framerule=0pt,
  language=Java,
  numbers=none,
  columns=fullflexible,
  showspaces=false,
  showtabs=false,
  breaklines=true,
  showstringspaces=false,
  breakatwhitespace=true,
  commentstyle=\color{pgreen},
  keywordstyle=\color{pblue},
  stringstyle=\color{pred},
  keywords={public, List, new, extends, private, void, for, if, else, throw, throws, null, return, Class, Method, Exception, static, Object},
  keywordstyle=[2]\color{level2},
  keywords=[2]{FOR, IF, ENDIF, ENDFOR, ELSE, compile},
  basicstyle={\small\ttfamily},
  moredelim=[il][\textcolor{pgrey}]{\$\$},
  moredelim=[is][\textcolor{pgrey}]{\%\%}{\%\%}
}

\lstdefinelanguage{contract}{
  xleftmargin=0.8em,
  framexleftmargin= 0.5em,
  frame=single,
  framerule=0pt,
  language=Java,
  columns=fullflexible,
  showspaces=false,
  showtabs=false,
  breaklines=true,
  showstringspaces=false,
  breakatwhitespace=true,
  commentstyle=\color{pgreen},
  keywordstyle=\color{pblue},
  stringstyle=\color{pred},
  keywords={Contract, definition, precondition, postcondition},
  keywordstyle=[2]\color{level2},
  keywords=[2]{String, Real, let, in, and, any, allInstance, oclIsNew, Integer, result, self, pre, includes, Boolean},
  basicstyle={\small\ttfamily},
  moredelim=[il][\textcolor{pgrey}]{\$\$},
  moredelim=[is][\textcolor{pgrey}]{\%\%}{\%\%}
}

\definecolor{level2}{RGB}{180,30,80}
\definecolor{level3}{RGB}{36,148,227}
\definecolor{level4}{RGB}{120,120,120}
\definecolor{level5}{RGB}{150,150,150}

\lstdefinelanguage{RE}{
keywordstyle=\color{},
keywordstyle=[2]\color{level2}\bfseries,
keywordstyle=[3]\color{level3}\mdseries,
keywordstyle=[4]\color{level4},
keywordstyle=[5]\color{level5},
basicstyle={\tiny\ttfamily},
numberstyle=\tiny\color{gray},
frame=b,
columns=fullflexible,
showstringspaces=false,
morekeywords={},
keywords=[2]{returns, name, def, pre, post, Integer, String, Real, definition, precondition, postcondition, and},
keywords=[3]{RequirementModel, Actor, Service, Entity, Contract, UC, extends, Operation, Refer, INV, TempProperty, Attribute, Reference},
keywords=[4]{in,out},
keywords=[5]{var}
}

\usepackage{fancyvrb,newverbs,xcolor}

\newverbcommand{\cverb}
  {\setbox\verbbox\hbox\bgroup}
  {\egroup\colorbox{cverbbg}{\box\verbbox}}

\definecolor{CR}{RGB}{0,69,122}
\definecolor{FGCR}{RGB}{253,177,26}
\definecolor{InvalidCR}{RGB}{175,32,67}
\definecolor{ValidCR}{RGB}{37,165,203}

\begin{document}
%
\title{Automated Prototype Generation from Formal Requirements Model}

\author{Yilong~Yang, 
        ~Xiaoshan~Li, 
        ~Zhiming~Liu,
        ~Wei~Ke,
        ~Quan~Zu,
        and~Xiaohong~Chen
\IEEEcompsocitemizethanks{




\IEEEcompsocthanksitem Yilong~Yang and Xiaoshan~Li are with the Faculty of Science and Technology, University of Macau, Macau.
\IEEEcompsocthanksitem Zhiming~Liu is with School of Computer and Information Science, Southwest University, Chongqing, China. 

\IEEEcompsocthanksitem Wei~Ke is with Macau Polytechnic Institute, Macau. 

\IEEEcompsocthanksitem Quan~Zu is with School of Software Engineering, Tongji University. Shanghai, China.

\IEEEcompsocthanksitem Xiaohong~Chen is with Department of Computer Science, University of Illinois at Urbana-Champaign, USA. 

\IEEEcompsocthanksitem Zhiming~Liu is the corresponding author of this paper. Email: zhimingliu88@swu.edu.cn.

}
}
\IEEEtitleabstractindextext{%
\begin{abstract}
\justifying
Prototyping is an effective and efficient way of requirement validation to avoid introducing errors in the early stage of software development. However, manually developing a prototype of a software system requires additional efforts, which would increase the overall cost of software development. In this paper, we present an approach with a developed tool to automatic generation of prototypes from formal requirements models. A requirements model consists of a use case diagram, a conceptual class diagram, use case definitions specified by system sequence diagrams and the contracts of their system operations. A system operation contract is formally specified by a pair of pre- and post-conditions in OCL. We propose a method to decompose a contract into executable parts and non-executable parts.  A set of transformation rules is given to decompose the executable part into pre-implemented primitive operations. A non-executable part is usually realized by significant algorithms such as sorting a list, finding the shortest path or domain-specific computation. It can be implemented manually or by using existing code. A CASE tool is developed that provides an interface for developers to develop a program for each non-executable part of a contract, and automatically transforms the executables into sequences of pre-implemented primitive operations. We have conducted four cases studies with over 50 use cases. The experimental result shows that the 93.65\% of requirement specifications are executable, and only 6.35\% are non-executable such as sorting and event-call, which can be implemented by developers manually or invoking the APIs of advanced algorithms in Java library. The one second generated the prototype of a case study requires approximate nine hours’ manual implementation by a skilled programmer. Overall, the result is satisfiable, and the proposed approach with the developed CASE tool can be applied to the software industry for requirements engineering.
\end{abstract}

}

\maketitle

\IEEEdisplaynontitleabstractindextext

%
\IEEEpeerreviewmaketitle

\IEEEraisesectionheading{\section{Introduction}\label{sec:introduction}}
\IEEEPARstart{R}{equirements} error is one of the causes leading failings in software projects \cite{DBLP:journals/re/SutcliffeEM99}. Careful requirements modeling along with systematic validation helps to reduce the uncertainty about target systems \cite{Hofmann2001}\cite{sommerville2015software}. The goal of requirements validation and evolution is to construct the consistent requirements for the needs of target users \cite{Atladottir2012}. However, this process is complicated, and it can be hard to produce a correct and complete requirements specification. The complexity is due to the following interrelated attributes: 1) the complexity of application domains and business processes; 2) the uncertainty of clients and domain experts about their needs; 3) the lack of the understanding of system developers about application domains; 4) the difficulties of the understanding between system developers and clients \cite{Lichter1994}. 



%
%
Rapid prototyping is an effective approach to requirements validation and evolution via an executable model of a software system to demonstrate concepts, discover requirements errors and find possible fixing solutions, and discover missing requirements \cite{Kordon2002}. Besides the implementation of main system functionalities, a prototype has a User Interface (UI) \cite{userinterface} that allows the client to validate their requirements visually, make it easy to find out faults in the requirements, and then fix them \cite{Kamalrudin2011}. In practice, it is very desirable to have a tool that generates prototypes directly from requirements automatically. However, state-of-the-art research and Computer-Aided Software Engineering (CASE) tools still have long distances to reach the goal \cite{Ciccozzi2018}. Unified Modeling Language (UML) is the de facto standard for requirement modeling and system design. Current UML modeling tools, such as Rational Rose, SmartDraw, MagicDraw, Papyrus UML, can only generate skeleton code, where classes only contain attributes and signatures of operations, not their implementations \cite{Regep2000}. Even when design models (e.g., sequence diagrams) are provided, only less than 48\% correct source code can be generated \cite{Kundu2013} by AndroMDA\footnote{\url{http://andromda.sourceforge.net}} and MasterCraft \cite{kulkarni2002generating} from design models.
 

\begin{figure*}[!htb]
  \centering
  \includegraphics[width=1\textwidth]{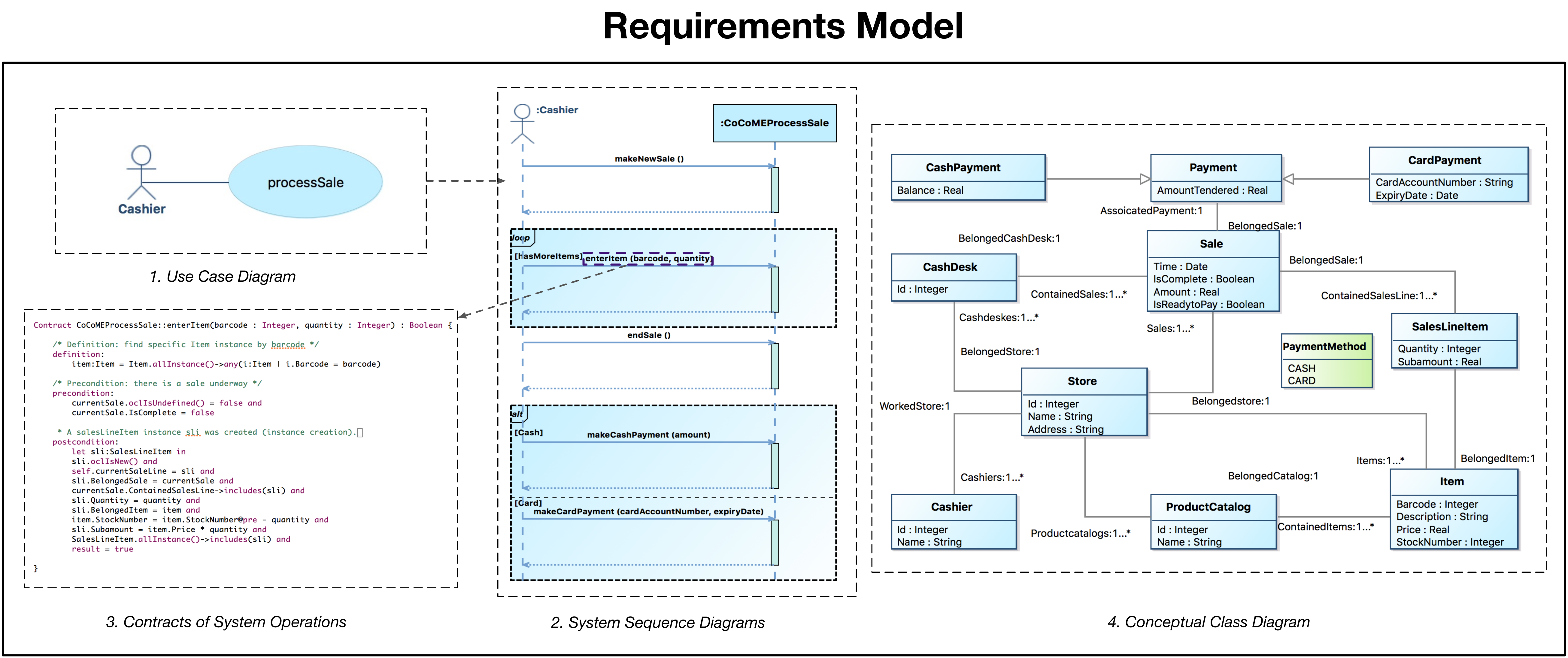}
  \caption{Requirements Model}\label{rm}
\end{figure*}

%
%
In this paper, we present an approach to automatically prototype generation from a requirements model in UML diagrams complemented by formal contracts of system operations. Compared with other related work, our approach does not require design models but rely on a requirements model \cite{DBLP:conf/compsac/LiLH01}\cite{zhiming2003}\cite{DBLP:journals/scp/ChenLRSZ09} in Figure \ref{rm}, which contains:

\noindent\textit{$\bullet$ A use case diagram:} 
A use case diagram captures domain processes as use cases in terms of interactions between the system and its users. It contains a set of use cases for a system, the actors represented a type of users of the system or external systems that the system interacts with, the relations between the actors and these use cases, and the relations among the use cases. It helps customers and domain experts specify functional requirements of the target system, and it assists in generating the operation list to hold system operation in prototypes, which is shown in Figure \ref{prototypesf}. 


\vspace{.1cm}
\noindent\textit{$\bullet$ System sequence diagrams.} A system sequence diagram describes a particular domain process of a use case. It contains the actors that interact with the system, the system and the system events that the actors generate, their order, and inter-system events. Compared with the sequence diagram, a system sequence diagram treats all systems as a black box and contains system events across the system boundary between actors and systems without object lifelines and internal interactions between objects. It helps customers to find system operations and provides the sequences to interact with the prototype for requirements validation.

\noindent\textit{$\bullet$ Contracts of system operations:}
The contract of a system operation \cite{liu2002object}\cite{meyer2002design} specifies the conditions that the state of the system is assumed to satisfy before the execution of the system operation, called the {\em pre-condition} and the conditions that the system state is required to satisfy after the execution (if it terminated), called the {\em post-condition} of the system operation. Typically, the pre-condition specifies the properties of the system state that need to be checked when system operation is to be executed, and the post-condition defines the possible changes that the execution of the system operation is to realize. A state of an object-oriented system is about the existing objects together with their properties/states and relations/links. The state is an object diagram defined by a conceptual class diagram plus the input and returns parameters of the operations. The changes of system states are classified into i) new objects created (together with initial values of attributes and links of associations), ii) attributes of existing objects (in the current state) modified, iii) new links among existing objects formed, and iv) existing objects and/or links are removed. The basic operations of changing the state are defined as the \textit{primitive operations} in our approach. We will see, the decomposition from system operation into primitive operations is the theoretical foundation of our approach to automatic generation of an abstract and yet executable model from a requirements model. 

\noindent\textit{$\bullet$ A conceptual class diagram.}  A conceptual class diagram illustrates abstract and meaningful concepts and their relations in the problem domain, in which the concepts are specified as classes, the relations of the concepts are specified as the associations between the classes, and the properties of the concepts are specified as the attributes of the classes. The proposed approach can directly generate to Java classes that represent domain concepts and encapsulates the primitive operations such as getting and setting the values of attributes, adding and remove links, and finding an object through links in the prototypes.

The idea of automated prototype generation from a requirements model is presented in our earlier work \cite{Li2008}\cite{Li2010}. There, the feasibility is demonstrated with a small example. In this paper, we extend original methods and propose new algorithms to 1) automated decomposition of the system operations into primitive operations, and encapsulation them into the classes based on object-oriented design patterns \cite{alexander1977pattern}\cite{vlissides1995design}\cite{richards2015software}\cite{Fowler2002}, 2) non-executable analysis of system operation contract, and wrap the non-executable parts of contract into a interface, it further can be implemented by developers and third-party APIs, 3) requirements validation and revolution based on the prototype mechanisms of object state observation and pre-condition and invariant checking.
The evaluation result from four case studies shows that our approach can correctly generate 93.65\% code from requirements models, the remaining non-executable 6.35\% requirements can be recognized and wrapped as an interface, which can be manually implemented or matched with third-party APIs libraries.

The remainder of this paper is organized as follows: Section 2 presents the preliminary of our approach. Section 3 introduces how to decompose a system operation into primitive operations. Section 4 presents how to generate the prototype. Section 5 provides how to use the generated prototype for requirement validation and refinement. Section 6 presents the experiments result of our approach on the four case studies of a Library Management System, CoCoME, ATM and a Loan Processing System. Section 7 and 8 discuss the related work, conclude this paper, and outline the future work.

\section{Preliminary}
In this section, we introduce the terminology used in the requirements model and prototypes.
%
%
\subsection{Terminology}
The terminology related to the proposed approach except the terminology introduced in the previous section are listed as follows:


\vspace{.1cm}
\noindent\textit{$\bullet$ Association and link.} An association is a relationship between two classes in a conceptual class diagram that specifies how instances of the classes can be linked to work together. A link is an instance of an association, which is a relationship between two objects in an object diagram.

\vspace{.1cm}
\noindent\textit{$\bullet$ Entity class and fabricated class.} To indicate the classes are from domain concepts or fabrications in the prototype, we divide classes into two type: entity classes are Java classes generated in prototypes from conceptual class diagrams, the others are fabricated classes. For example, Java class \textit{Item} is entity class generated from the conceptual class diagram in CoCoME, and Java class \textit{EntityManager} is a fabricated class that helps to find, create, and release the objects in the system.

\vspace{.1cm}
\noindent\textit{$\bullet$ Object reference and reference list.} In object-oriented programming such as Java, the value of a variable which has a type of a class is an object reference, which provides a way to access an object in the heap of a system. A reference list is a list of object references to access objects with the same type of a class. We will see, reference lists are used to record and access the objects of the prototype.

\vspace{.1cm}
\noindent\textit{$\bullet$ System operation.} System operation is an operation that the system executes in response to a system input event in system sequence diagrams.

\vspace{.1cm}
\noindent\textit{$\bullet$ Primitive operation.} Primitive operations are the operations introduced in our approach to covers all actions to manipulate objects, the attributes of objects, and the links of objects in Table 1. The details are shown in Section 3.1. 

\subsection{Object Constraint Language}
Object Constraint Language\footnote{\url{http://www.omg.org/spec/OCL/}} (OCL) is a part of UML. It is used mainly for specifying constraints of UML diagrams, such as pre and post conditions of operations and invariants. In this paper, we adopt OCL in the latest version \textit{v2.4} to specify the contracts of system operations. It not only can specify the pre-condition and post-condition of system operation but also allows to specify shared specifications from pre-condition and post-condition in the \textit{definition} section. The following example shows how OCL specify the contract of the system operation \textit{enterItem} (a cashier scans products in a sale process of a supermarket) of the use case \textit{processSale}.
\begin{lstlisting}[language=contract]
//Signature
Contract CoCoMEProcessSale::enterItem
  (barcode : String, quantity : Real) : Boolean {

//Definition Section
definition:
    //Find Object
    item:Item = Item.allInstance()->any(i:Item | i.Barcode = barcode)

//Pre-condition Section
precondition:
    currentSale.oclIsUndefined() = false and
    currentSale.IsComplete = false and
    item.oclIsUndefined() = false and
    item.StockNumber > 0

//Post-condition Section
postcondition:
    //Create an Object
    let sli:SalesLineItem in
    sli.oclIsNew() and
    //Add Links
    self.currentSaleLine = sli and
    sli.BelongedSale = currentSale and
    currentSale.ContainedSalesLine->includes(sli) and
    sli.BelongedItem = item and
    //Modify Attributes
    sli.Quantity = quantity and
    sli.Subamount = item.Price * quantity and
    item.StockNumber = item.StockNumber@pre - quantity and
    //Add an Object
    SalesLineItem.allInstance()->includes(sli) and
    result = true
}
\end{lstlisting}

\noindent\textbf{Signature:} The contract first specifies the signature of system operation \textit{enterItem()} of use case \textit{processSale}. The signature declares a \textit{String} variable \textit{barcode} and a \textit{Real} variable \textit{quantity} as input, and output variable typed \textit{Boolean}. 

\vspace{.2cm}

\noindent\textbf{Definition Section:} In the \textit{definition} section, we find the object \textit{item} of the class \textit{Item} with the attribute \textit{Barcode} equal to the input variable \textit{barcode}. We will see that the object \textit{item} is used in both the pre- and post-conditions. 

\vspace{.2cm}

\noindent\textbf{Pre-condition Section:} The pre-condition of \textit{enterItem} specifies that the current sale object \textit{currentSale} is \textit{existed}, and the value of attribute \textit{IsComplete} of \textit{currentSale} is equal to \textit{false}, the object \textit{item} with the scanned \textit{barcode} is \textit{existed} in the system, and the stock number is greater than zero. 

\vspace{.2cm}

\noindent\textbf{Post-condition Section:} The post-condition of \textit{enterItem} specifies that 1) a new object \textit{sli} of class \textit{SalesLineItem} was created, 2) the \textit{currentSaleLine} was linked to the new created object \textit{sli}, 3) the links among \textit{currentSale}, \textit{sli} and \textit{item} were formed, 4) the attributes \textit{Quantity} and \textit{Subamount} of \textit{sli} were set to the value of input variable \textit{quantity} and \textit{item.Price*quantity}, 5) the attribute \textit{StockNumber} of \textit{item} was reduced by the number of \textit{quantity}, 6) the new created object \textit{sli} was added in the object list \textit{SaleLineItem}, 7) the output of system operation \textit{enterItem()} was \textit{true}.

Note that system operations may manipulate the same objects in a system sequence diagram of a use case. OCL allows to access the new created object in the same use case. For example, the object \textit{currentSale} of the class \textit{Sale} is created by system operation \textit{makeNewSale()} of the use case \textit{processSale}. It can be reused in the contract of the system operation \textit{enterItem()}.



%
%
\section{System Operation Decomposition}
In this section, we first present a collection of primitive object-oriented operations and then introduce transformation rules and algorithms that automatically decompose a system operation to primitive operations. Finally, we present an example to show how the transformation rules and algorithms work. 



%
%
\subsection{Primitive Operations}
Referring to atomic actions for manipulation tables in relational databases, we introduce a collection of primitive object-oriented operations of the object-oriented system for system operation decomposition in Table \ref{atomicresponsibilitiy}, 
\begin{table}[!htb]
  \centering
  \ra{1.5}
    \caption{Primitive Operations}\label{atomicresponsibilitiy}
  \begin{adjustbox}{max width=0.48\textwidth}
  \begin{tabular}{cll @{}}
    \toprule
    & Primitive Operation & Return Type \\
    \midrule
     
    \multirow{5}{*}{Object} &\emph{\textbf{findObject}(ClassName$\mathord{:}$String, condition$\mathord{:}$String)} & Object  \\
    &\emph{\textbf{findObjects}(ClassName$\mathord{:}$String, condition$\mathord{:}$String)} & Set(Object) \\
    &\emph{\textbf{createObject}(ClassName$\mathord{:}$String)} & Object \\
    &\emph{\textbf{addObject}(ClassName$\mathord{:}$String, ob$\mathord{:}$Class)} & Boolean  \\
    &\emph{\textbf{releaseObject}(ClassName$\mathord{:}$String, ob$\mathord{:}$Class)} & Boolean \\
   
    \midrule
    \multirow{2}{*}{Attribute}&\emph{\textbf{getAttribute}(ob$\mathord{:}$Class, attriName$\mathord{:}$String)} & PrimeType \\
    &\emph{\textbf{setAttribute}(ob$\mathord{:}$Class, attriName$\mathord{:}$String, mathExp$\mathord{:}$String)} & Boolean\\

    \midrule
    \multirow{6}{*}{Link}&\emph{\textbf{findLinkedObject}(o$\mathord{:}$Class, assoName$\mathord{:}$String, condition$\mathord{:}$String)} & Object \\
    &\emph{\textbf{findLinkedObjects}(o$\mathord{:}$Class, assoName$\mathord{:}$String, condition$\mathord{:}$String)} & Set(Object)\\ 
    &\emph{\textbf{addLinkOnetoMany}(ob$\mathord{:}$Class, assoName$\mathord{:}$String, addOb$\mathord{:}$Class)} & Boolean \\
    &\emph{\textbf{addLinkOnetoOne}(ob$\mathord{:}$Class, assoName$\mathord{:}$String, addOb$\mathord{:}$Class)} & Boolean \\
    &\emph{\textbf{removeLinkOnetoMany}(ob$\mathord{:}$Class, assoName$\mathord{:}$String, removeOb$\mathord{:}$Class)} & Boolean\\
    &\emph{\textbf{removeLinkOnetoOne}(ob$\mathord{:}$Class, assoName$\mathord{:}$String)} & Boolean\\

    \bottomrule
  \end{tabular}
  \end{adjustbox}
\end{table}
which covers all manipulations to a) objects, b) the attributes of objects, and c) the links of objects. 

\vspace{.2cm}
\noindent\textbf{Objects:} The following primitive operations are used to manipulate objects. An object or objects can be retrieved through primitive operation \textit{findObject()} or \textit{findObjects()} with a class name and a query condition. An object can be created by primitive operation \textit{createObject()} with a class name, and then the created object can be added to the system through primitive operation \textit{addObject()} with providing a class name and an object reference \textit{ob}. Primitive operation \textit{releaseObject()} can be used to delete an object from the system by providing a class name and an object reference \textit{ob}. 

\vspace{.2cm}
\noindent\textbf{Attributes of Objects:} The next two primitive operations are used for getting and setting the value of an attribute. Primitive operation \textit{getAttribute()} can retrieve the value of an attribute of an object by providing an object reference \textit{ob} and the name of attribute. The value of an attribute of an object can be changed by the primitive operation \textit{setAttribute()} with an object reference \textit{ob}, the name \textit{attriName} of an attribute, and a math expression \textit{mathExp}. 

\vspace{.2cm}
\noindent\textbf{Links of Objects:} The links can be used to find the linked objects. Note that we use different primitive operations to manipulate links corresponding with two different types of the association of classes (one-to-one and one-to-many relation). By providing an object reference \textit{ob}, an association name \textit{assoName}, and an condition, primitive operations \textit{findLinkedObject()} and \textit{findLinkedObjects()} can retrieve the linked object or objects. An link from object \textit{ob} to object \textit{addOb} can be formed by invoking primitive operation \textit{addLinkOnetoMany()} or \textit{addLinkOnetoOne()} with the name of the association, and the object reference \textit{ob} and \textit{addOb}. Primitive operation \textit{removeLinkOnetoMany()} or \textit{removeLinkOnetoOne()} can be used to break the link by providing an object reference \textit{ob}, and the name of the association \textit{assoName}. Primitive operation \textit{removeLinkOnetoMany()} requires providing a reference to the target object \textit{removeOb}, and the reference indicates which link will be removed from the object. 
%


%
%
\subsection{Transformation Rules}
We have introduced the contract of system operation and primitive operations. In this subsection, we present how to transform an OCL contract to primitive operations. Transformation rules will be presented in this form:
\[
\begin{array}{l}
    \textit{Rule}: \frac{\textit{OCL Expression}}{\textit{Primitive Operation in Java Code}}
\end{array}
\]
The transformation rule contains two parts: the above section is an OCL Expression, and the bottom part is a primitive operation in Java code. The transformation rules form OCL expressions and primitive operations as pairs, and those pairs are the foundation of the transformation algorithm for automatic system operation decomposition. In short, we refine the original ten transformation rules in the previous work \cite{Li2008}\cite{Li2010} to twenty-five transformation rules that cover all the primitive operations corresponding in the contract sections of definition, pre-condition, and post-condition. The difference from our previous work is discussed in the related work section. Follow the convention of OCL contract, we present those transformation rules into three parts: a) definition transformation, b) pre-condition transformation, and c) post-condition transformation. 

\subsubsection{Definition Section Transformation}

The definition section of the contract specifies that the objects are further used in pre-condition and post-condition. In object-oriented system, objects can be reached through the links of objects, which are defined by the associations of the classes. In our approach, we build pure fabricated class \textit{EntityManager} (EM) to record all the references of objects in the system. Therefore, objects can be found through \textit{EntityManager} with a query condition, and then other related objects can be reached through the links of the founded objects. In definition section, seven transformation rules are presented, which involve the primitive operations \textit{findObject()}, \textit{findObjects()}, \textit{findAssociationObject()}, and \textit{findAssociationObjects()} of Table \ref{atomicresponsibilitiy}.
\[
\begin{array}{l}
    \mathbf{R_1}: \frac{\textit{obs$\mathord{:}$\textbf{Set}(ClassName)=ClassName.\textbf{allInstances}()}}{\emph{List<ClassName> obs = EM.\textbf{findObjects}(ClassName$\mathord{:}$String)}}
\end{array}
\]
The rule $R_1$ shows finding all the objects \textit{obs} of the class named \textit{ClassName} in the system. This OCL expression is mapped to the primitive operation \textit{findObjects()}, and the found objects are assigned to a list reference \textit{obs} of the class \textit{ClassName}.
\[
\footnotesize
\begin{array}{l}
   \mathbf{R_2}: \frac{\textit{obs$\mathord{:}$\textbf{Set}(ClassName)=ClassName.\textbf{allInstances}()$\rightarrow$\textbf{select}(o $|$ \textbf{conditions}(o))}}{\emph{List<ClassName> obs = EM.\textbf{findObjects}(ClassName$\mathord{:}$String, conditions(o)$\mathord{:}$String)}}
\end{array}
\]
\[
\footnotesize
\begin{array}{l}
    \mathbf{R_3}: \frac{\textit{ob$\mathord{:}$ClassName = ClassName.\textbf{allInstances}()$\rightarrow$\textbf{any}(o $|$ \textbf{conditions}(o))}}{\textit{ClassName ob = EM.\textbf{findObject}(ClassName$\mathord{:}$String, conditions(o)$\mathord{:}$String)}}
\end{array}
\]
Based on rule $R_1$, the rules $R_2$ and $R_3$ are introduced to find objects \textit{obs} or an object \textit{ob} from all the instances of the class named \textit{ClassName} with the constraints \textit{condition(o)} by using \textit{select} or \textit{any} OCL keywords. \textit{condition(o)} is a logic formula about object \textit{o}, which composites atomic formulae of pre-condition with the logical operators \textit{and} and \textit{or}. The OCL expressions of $R_2$ and $R_3$ are mapped to the primitive operation \textit{findObjects()} or \textit{findObject()}, and then assign the found objects or object to a reference list \textit{obs} or a reference \textit{ob} of class \textit{ClassName}. 

For example, when a cashier scans the product by invoking the system operation \textit{enterItem()}, the system will find the object \textit{item} with the specific \textit{barcode}, which has the same value retrieved from the scanner. This part functional semantics are specified in the definition section of contract \textit{enterItem()}. It will map to the primitive operation \textit{findObject()} by $R_3$ as follows:
\[
\begin{array}{l}
    \frac{\textit{item$\mathord{:}$Item = Item.\textbf{allInstances}()$\rightarrow$\textbf{any}(o $|$ i.Barcode = barcode)}}{\textit{Item ob = EM.\textbf{findObject}("Item", "i.Barcode = barcode")}}
\end{array}
\]
Note that the condition \textit{i.Barcode = barcode} will be further mapped by $R_{12}$. Once we find the target objects, the related objects can be found through the links of the target objects. Those transformations present in the next four rules:

\[
\footnotesize
\begin{array}{l}
    \mathbf{R_4}: \frac{\textit{o$\mathord{:}$ClassName = ob.assoName}}{\emph{ClassName o =  EM.\textbf{findLinkedObject}(ob$\mathord{:}$Class, assoName$\mathord{:}$String)}}
\end{array}
\]
\[
\footnotesize
\begin{array}{l}
    \mathbf{R_5}: \frac{\textit{obs$\mathord{:}$Set(ClassName) = ob.assoName}}{\emph{List<ClassName> obs = EM.\textbf{findLinkedObjects}(ob$\mathord{:}$Class, assoName$\mathord{:}$String)}}
\end{array}
\]
The rules $R_4$ and $R_5$ show finding the linked object through \textit{ob.assoName}, where the association may be one-to-one or one-to-many relationship. If \textit{assoName} is one-to-one association, \textit{ob.assoName} will return a object reference \textit{o} to the object linked with object \textit{ob}; otherwise \textit{ob.assoName} returns a reference list \textit{obs}. Therefore, the OCL expressions of $R_4$ and $R_5$ are mapped to primitive operations \textit{findLinkedObject()} and \textit{findLinkedObjects()} with the input variables: an object reference \textit{ob} and an association name \textit{assoName} of object \textit{ob}, and then assign the found object or object list to the reference \textit{o} or reference list \textit{obs} correspondingly.
\[
\scriptsize
\begin{array}{l}
    \mathbf{R_6}: \frac{\textit{obs$\mathord{:}$Set(ClassName) = ob.assoName$\rightarrow$\textbf{select}(o $|$ \textbf{conditions}(o))}}{\emph{List<ClassName> obs = EM.\textbf{findLinkedObjects}(ob$\mathord{:}$Class, assoName$\mathord{:}$String, preconditions(o)$\mathord{:}$String)}}
\end{array}
\]
\[
\footnotesize
\begin{array}{l}
    \mathbf{R_7}: \frac{\textit{o$\mathord{:}$ClassName = ob.assoName$\rightarrow$\textbf{any}(o $|$ \textbf{conditions}(o))}}{\emph{ClassName ob =  EM.\textbf{findLinkedObject}(ob$\mathord{:}$Class, assoName$\mathord{:}$String, conditions(o)$\mathord{:}$String)}}
\end{array}
\]
Based on the rule $R_5$, the rules $R_6$ and $R_7$ apply OCL expression \textit{select} and \textit{any} with a query condition \textit{condition(o)} to \textit{ob.assoName} with navigation $\rightarrow$ to filter the specific objects from the associated objects. 

For example, we find a \textit{SaleLineItem} object under current sale \textit{s} with quantity number great than 2. The corresponding OCL contract and $R_7$ transformation are:
\[
\small
\begin{array}{l}
    \frac{\textit{sli$\mathord{:}$SaleLineItem = s.ContainedSaleLine$\rightarrow$\textbf{any}(sl $|$ sl.Quantity > 2)}}{\emph{SaleLineItem sli =  EM.\textbf{findLinkedObject}(s, "ContainedSaleLine", "sl.Quantity > 2")}}
\end{array}
\]
The same as the previous example, the condition \textit{sl.Quantity > 2} will make a further transformation by rule $R_{12}$.


%
%
\subsubsection{Pre-condition Transformation}
The pre-condition section of the contract specifies the status of the objects before execution of system operation. The related objects have been specified in the definition section of the contract. The pre-condition section specifies the constraints on those objects before execution of system operation. The pre-condition transformation maps those constraints to the primitive operations, which involves \textit{getAttribute()} and a set of basic checking operations under the fabricated class \textit{StandardOPs}. In short, eight transformation rules from $R_8$ to $R_{15}$ are presented in the pre-condition section.
\[
\begin{array}{l}
    \mathbf{R_8}: \frac{\textit{ob.oclIsUndefined() = bool}}{\emph{\textbf{StandardOPs.oclIsUndefined}(ob$\mathord{:}$Class, bool$\mathord{:}$Boolean)}}
\end{array}
\]
\[
\begin{array}{l}
    \mathbf{R_9}: \frac{\textit{var.oclIsTypeOf(type)}}{ \emph{\textbf{StandardOPs.oclIsTypeOf}(\flqq var\frqq, type$\mathord{:}$String)}}
\end{array}
\]
\[
\begin{array}{l}
    \mathbf{R_{10}}: \frac{\textit{obs.isEmpty() = bool}}{\emph{\textbf{StandardOPs.isEmpty}(obs$\mathord{:}$Set(Class), bool$\mathord{:}$Boolean)}}
\end{array}
\]

The rules $R_8$, $R_9$, and $R_{10}$ map the basic OCL checking expression about object and object list to the primitive operations. The contract of $R_8$ checks that the reference \textit{ob} does not refer to an object. The OCL expression of $R_9$ is used to check that the variable \textit{var} conforms the specific \textit{type}, in which the \textit{var} is a variable of primitive type, an object reference, or a reference list. The contract of $R_{10}$ is to check the object list \textit{obs} is empty.
\[
\begin{array}{l}
    \mathbf{R_{11}}: \frac{\textit{obs.size() op\tnote{1} \textit{ }mathExp}}{  \emph{\textbf{StandardOPs.size}(obs$\mathord{:}$Set(Class)) \flqq op\frqq \textit{ }\flqq mathExp\frqq}}
\end{array}
\]
\[
\begin{array}{l}
    \mathbf{R_{12}}: \frac{\textit{ob.AttriName op varPM\tnote{6}}}{\emph{\textbf{getAttribute}(ob$\mathord{:}$Class, attriName$\mathord{:}$String) \flqq op\frqq\textit{ }\flqq varPM\frqq}}
\end{array}
\] 
The rules $R_{11}$ and $R_{12}$ map the logic expression from OCL to Java. The \textit{op} is an infix comparison operator in OCL. \flqq op\frqq\textit{ }represents the corresponding Java code of \textit{op} after compilation. Most operators of OCL and Java are the same, except equal operator \textit{=} and non-equal \textit{<>} of OCL. Those operators will be compiled to Java operators \textit{==} and \textit{!=}. Furthermore, when we compare between two variables \textit{s1} and \textit{s2} of \textit{String} type, the above operators will be compiled to Java code \textit{s1.equals(s2)} and \textit{!s1.equals(s2)} respectively. The rule $R_{11}$ gets the size of list, and then checks the constraint of the list size against a math expression \textit{mathExp}. The \textit{mathExp} may contain numbers, variables, operators, functions and brackets. The rule $R_{12}$ firstly retrieves the value of the attribute \textit{attriName} of the object \textit{ob} by transforming \textit{ob.attribute} to the primitive operation \textit{getAttribute}, and checks the constraint on the value of the attribute through the expression \textit{op} \textit{varPM}. The \textit{varPM} is a variable of primitive type or a math expression. 

For example, the pre-condition \textit{item.StockNunmber > 0} of contract \textit{enterItem}. ">" is the \textit{op} expression. "0" is the \textit{varPM} expression. This pre-condition will be compiled to Java code \textit{item.getStockNumber() > 0} by rule $R_{12}$:
\[
\begin{array}{l}
    \frac{\textit{sl.Quantity > 2}}{\emph{\textbf{getAttribute}(sl, "Quantity") > \textit{ }2}}
\end{array}
\] 
The rules $R_{13}$ and $R_{14}$ use the OCL expression \textit{ClassName.allInstances()} of the definition section to find all objects of class \textit{ClassName}, and to check whether the specific object \textit{ob} is included in or excluded from the object list. Therefore, \textit{ClassName.allInstances()} is mapped to \textit{EM.findObjects(ClassName)}, and then the founded object will be passed to the operations \textit{includes()} or \textit{excludes()} of the class \textit{StandardOPs}.
\[
\begin{array}{l}
    \mathbf{R_{13}}: \frac{\textit{ClassName.\textbf{allInstances}()$\rightarrow$\textbf{includes}(ob)}}{\emph{\textbf{StandardOPs.includes}(EM.findObjects(ClassName), ob$\mathord{:}$Class)}}
\end{array}
\] 
\[
\begin{array}{l}
    \mathbf{R_{14}}: \frac{\textit{ClassName.\textbf{allInstances}()$\rightarrow$\textbf{excludes}(ob)}}{\emph{\textbf{StandardOPs.excludes}(EM.findObjects(ClassName), ob$\mathord{:}$Class)}}
\end{array}
\]
\[
\begin{array}{l}
    \mathbf{R_{15}}: \frac{\textit{ClassName.\textbf{allInstance}()->\textbf{isUnique}(o$\mathord{:}$ClassName | o.AttriName)}}{\emph{\textbf{StandardOPs.isUnique}(ClassName$\mathord{:}$String, AttriName$\mathord{:}$String)}}
\end{array}
\]     
The last rule $R_{15}$ of precondition section maps the unique detection expression to the operation \textit{isUnique()} of class \textit{StandardOPs}. This opeartion will get all the objects of class \textit{ClassName}, and check the specific attribute \textit{AttriName} has the unique value or not. Note that 1) OCL expressions of pre-condition section can also be used to specify the invariants of the system. Therefore, the transformation rules related pre-condition can also apply to the invariants transformation. 2) The implementation of class \textit{StandardOPs} is not presented in this paper, the details can be found on the GitHub repository\footnote{\url{https://github.com/RM2PT/CaseStudies/wiki/StandardOP}}.


%
%
\subsubsection{Post-condition Transformation}
The post-condition section of the contract specifies the status of objects after execution of the operation. Concretely, the post-condition specifies that the object was created, added, and released, the association was formed and broken, the state of attribute was updated. The related primitive operations are \textit{createObject()}, \textit{addObject()}, \textit{releaseObject()}, \textit{addOnetoManyAssociation()}, \textit{addOnetoOneAssociation()}, \textit{removeOnetoManyAssociation()}, \textit{removeOnetoOneAssociation()}, and \textit{setAttribute()}. Ten transformation rules of the post-condition are presented in post-condition section. The first rule $R_{16}$ is
\[
\begin{array}{l}
    \mathbf{R_{16}}: \frac{\textit{\textbf{let} ob$\mathord{:}$ClassName \textbf{in} ob.\textbf{oclIsNew}()}}{\emph{ClassName ob = EM.\textbf{createObject}(ClassName$\mathord{:}$String)}}
\end{array}
\]
The OCL expression \textit{let..in} of the rule $R_{16}$ describes the scope of object \textit{ob}. The \textit{ob.oclIsNew()} specifies that the object \textit{ob} was created after the execution of system operation. In order to make system into this state, the rule $R_{16}$ maps \textit{let..in} and \textit{oclIsNew()} to the primitive operation \textit{createObject()}, and then assigns the created object to the reference \textit{ob} of the class named as \textit{ClassName}. For example, the post-condition of system operation \textit{enterItem()} specifies that object \textit{sli} of class \textit{SaleLineItem} was created. By applying the rule $R_{16}$, this OCL expressions are mapped to:
\[
\begin{array}{l}
    \frac{\textit{\textbf{let} sli$\mathord{:}$SaleLineItem \textbf{in} sli.\textbf{oclIsNew}()}}{\emph{SaleLineItem sli = EM.\textbf{createObject}("SaleLineItem")}}
\end{array}
\]
The OCL expression of the rules $R_{17}$ and $R_{18}$ have already been used in the pre-condition to check whether the instance list of class \textit{ClassName} includes or excludes the object \textit{ob} or not. Those OCL expressions in the post-condition indicate that the object list of class \textit{ClassName} includes and excludes the object \textit{ob} after the execution of the system operation. 
\[
\begin{array}{l}
    \mathbf{R_{17}}: \frac{\textit{ClassName.\textbf{allInstances}()$\rightarrow$\textbf{includes}(ob)}}{EM.\emph{\textbf{addObject}(ClassName$\mathord{:}$String, ob$\mathord{:}$Class)}}
\end{array}
\] 
\[
\begin{array}{l}
    \mathbf{R_{18}}: \frac{\textit{ClassName.\textbf{allInstances}()$\rightarrow$\textbf{excludes}(ob)}}{EM.\emph{\textbf{releaseObject}(ClassName$\mathord{:}$String, ob$\mathord{:}$Class)}}
\end{array}
\] 
It is necessary to add or delete the object \textit{ob} to or from the object list to make the system status conforming the post-condition. Therefore, the \textit{includes()} and \textit{excludes()} with \textit{allInstances()} expression in the post-condition are mapped to the primitive operations \textit{addOjbect()} and \textit{releaseObject()}. 
\[
\begin{array}{l}
    \frac{\textit{SalesLineItem.\textbf{allInstances}()$\rightarrow$\textbf{includes}(sli)}}{EM.\emph{\textbf{addObject}("SalesLineItem", "sli")}}
\end{array}
\] 
For example, if we apply the rule $R_{17}$ to post-condition of system operation \textit{enterItem()}, the post-condition specifies that the created object \textit{sli} was included in object list of class \textit{SalesLineItem} will be mapped to primitive operation \textit{addObject()} with parameter class name \textit{SalesLineItem} and object name \textit{sli}.
\[
\begin{array}{l}
    \mathbf{R_{19}}: \frac{\textit{ob.assoName$\rightarrow$\textbf{includes}(addOb)}}{\emph{\textbf{addLinkOnetoMany}(ob$\mathord{:}$Class, assoName$\mathord{:}$String, addOb$\mathord{:}$Class)}}
\end{array}
\]     
\[
\begin{array}{l}
    \mathbf{R_{20}}: \frac{\textit{ob.assoName$\rightarrow$\textbf{excludes}(removeOb)}}{\emph{\textbf{removeLinkOnetoMany}(ob$\mathord{:}$Class, assoName$\mathord{:}$String, removeOb$\mathord{:}$Class)}}
\end{array}
\]
OCL expression \textit{includes} and \textit{excludes} can also be applied for the association \textit{ob.assoName}. That means the link has been formed or broken after executing the system operation. Therefore, the above rules $R_{19}$ and $R_{20}$ map those expressions to the primitive operations \textit{addLinkOnetoOne()} and \textit{removeLinkOnetoMany()}. 
\[
\begin{array}{l}
    \mathbf{R_{21}}: \frac{\textit{ob.assoName = addOb}}{\emph{\textbf{addLinkOnetoOne}(ob$\mathord{:}$Class, assoName$\mathord{:}$String, addOb$\mathord{:}$Class)}}
\end{array}
\]
The one-to-one association is implemented as an attribute of the class. The rule $R_{21}$ describes if the post-condition specifies that the one-to-one association \textit{ob.assoName} has an linked object \textit{addOb}. The primitive operation \textit{addLinkOnetoOne()} will be executed to make system to satisfy this post-condition. For examples, the links between current sale and the created object \textit{SalesLineItem} object were formed in following transformations:
\[
\begin{array}{l}
    \frac{\textit{currentSale.ContainedSalesLine$\rightarrow$\textbf{includes}(sli)}}{\emph{\textbf{addLinkOnetoMany}(currentSale, "ContainedSalesLine", sli)}}
\end{array}
\]    
\[
\begin{array}{l}
    \frac{\textit{sli.BelongedSale = currentSale}}{\emph{\textbf{addLinkOnetoOne}(sli, "BelongedSale", currentSale)}}
\end{array}
\]
Note that the association from class \textit{Sale} to \textit{SalesLineItem} is multiple, and the reversed association is a one-to-one association. Therefore, the rule $R_{19}$ will be triggered for transformating OCL expression to primitive operation \textit{addLinkOneToMany()} with an object \textit{currentSale}, the association name \textit{ContainedSalesLine}, and the created object \textit{sli} of class  \textit{SalesLineItem}. The rule $R_{21}$ is trigger for transformation OCL expression to primitive operation \textit{addLinkOneToOne()} with the object reference \textit{sli} of class \textit{SalesLineItem}, the association name \textit{BelongedSale}, and the reference \textit{currentSale} to current sale object.
\[
\begin{array}{l}
    \mathbf{R_{22}}: \frac{\textit{ob.assoName = \textbf{null}}}{\emph{\textbf{removeLinkOnetoOne}(ob$\mathord{:}$Class, assoName$\mathord{:}$String)}}
\end{array}
\]
The rule $R_{22}$ presents if the post-condition specifies that one-to-one association \textit{ob.assoName} has no associated object, the primitive operation \textit{removeLinkOnetoOne()} will be executed to broke the link that makes the system state conforming this post-condition. 
\[
\begin{array}{l}
    \mathbf{R_{23}}: \frac{\textit{ob.attriName = \textit{mathExp}}}{\emph{\textbf{setAttribute}(ob$\mathord{:}$Class, attriName$\mathord{:}$String, \flqq mathExp\frqq$\mathord{:}$PrimeType)}}
\end{array}
\]
The rule $R_{23}$ indicates that the attribute of the object \textit{ob.attriName} is equal to the math expression \textit{mathExp} in the post-condition. The primitive operation \textit{setAttribute} should be executed to conform this condition. 
For example, post-condition of \textit{enterItem()} specifies the sub-amount of \textit{SalesLineItem} is computed from the price of the item and the quantity of the item. 
\[
\begin{array}{l}
    \frac{\textit{sli.Subamount = item.Price * quantity}}{\emph{\textbf{setAttribute}(sli, "Subamount", \textbf{getAttribute}(item, "Price")*quantity)}}
\end{array}
\]
This OCL expression is mapped to primitive operations \textit{setAttribute()} and  \textit{getAttribute()}. Primitive operation \textit{getAttribute()} is used to get the price of the item, then the price is multiplied with quantity. Finally, the evaluated result is set to the attribute \textit{Subamount} of object \textit{SalesLineItem} named \textit{sli}.
\[
\footnotesize
\begin{array}{l}
    \mathbf{R_{24}}: \frac{\textit{obs->\textbf{forAll}(o$\mathord{:}$ClassName |
			o.AttriName = mathExp)}}{\textit{\textbf{for} (ClassName o$\mathord{ : }$obs) \{\textbf{setAttribute}(o$\mathord{:}$Class, AttriName$\mathord{:}$String, \flqq mathExp\frqq$\mathord{:}$PrimeType);\}}}
\end{array}
\]
We can use \textit{forAll} to literately specify the state of the objects in the post-condition. The rule $R_{24}$ extends the rule $R_{23}$ from specifying the state of a single object to a list of objects, that maps the OCL expression \textit{forAll} to Java \textit{for} loop expression.
\[
\begin{array}{l}
    \mathbf{R_{25}}: \frac{\textit{\textbf{return} = var\tnote{5}}}{\textit{\textbf{return} \flqq var\frqq;}}
\end{array}
\]
The rule $R_{25}$ specifies the state of return variables after system operation execution. The return variable \textit{var} can be a variable of prime type, a reference to a instances of class, or to a instance list of class. It will be mapped to the Java code correspondingly.
\[
\begin{array}{l}
    \mathbf{R_{26}}: \frac{\textit{\textbf{ThirdPartyServices}.opName(vars)}}
    {\textit{\textbf{service}.opName(\flqq vars\frqq)}}
\end{array}
\]
The rule $R_{26}$ is a special transformation rule that specifies the third-party APIs such as \textit{cardPayment()} and \textit{sorting()} used in the post-condition. This rule makes our approach extensible to invoking the third-party API library. It will transform the operation \textit{opName} and parameters \textit{vars} into the operation invoking in Java code.

\begin{figure*}[!htb]
  \centering
  \includegraphics[width=1\textwidth]{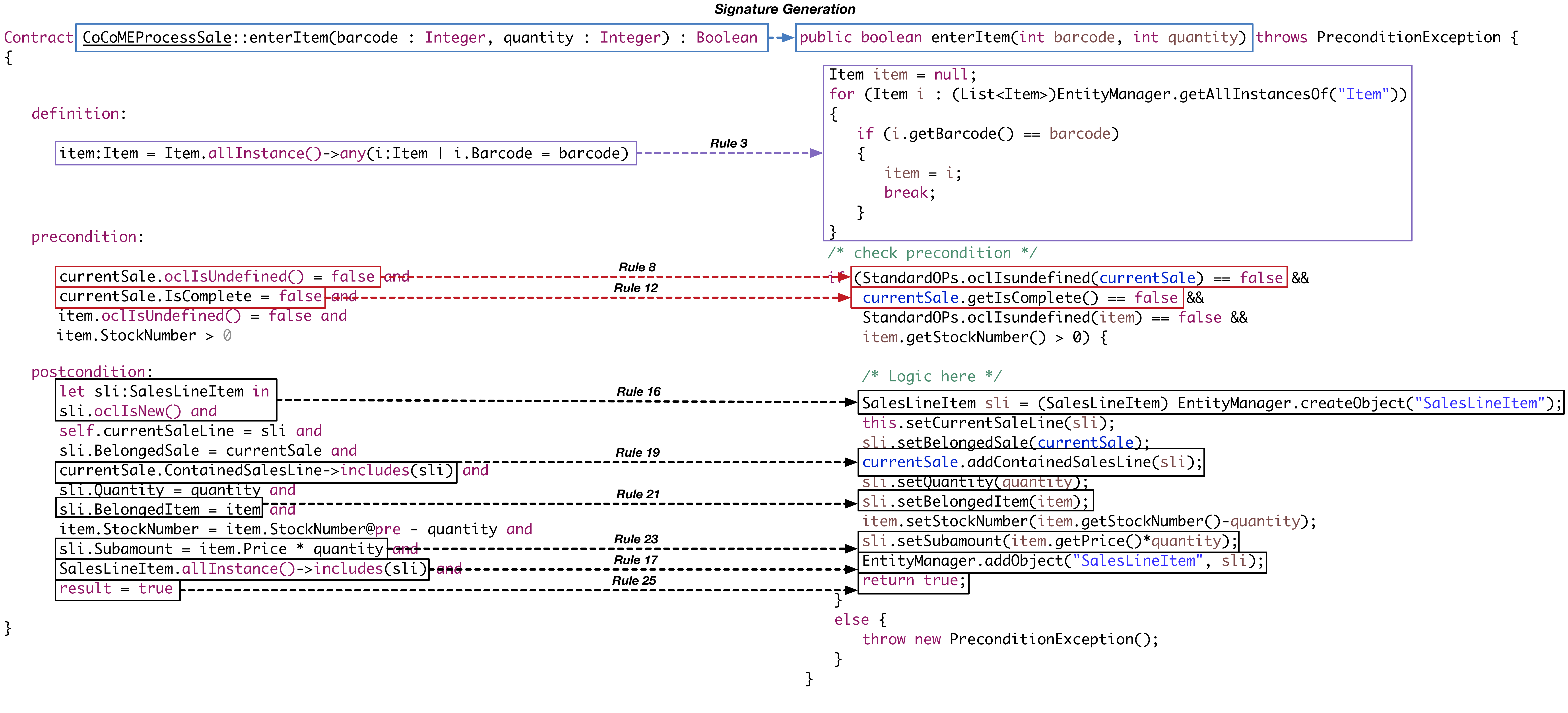}
  \caption{An Transformation Example for the Contract of \textit{enterItem()}}\label{mapEnterItem}
\end{figure*}
%
%
\subsection{Transformation Algorithm}
We have already presented all transformation rules for system operation decomposition. It is time to introduce the transformation algorithm that transforms the contract of system operation into primitive operations. Transformation algorithm is presented in Algorithm \ref{MappingAlgorithm}. It requires an OCL expression and a tag as input parameters, and return the mapped primitive operations. Note that the tag marks which section of the contract OCL expression comes from. The details of transformation algorithm are that initializing the result set \textit{rs} as empty set and index \textit{i} to zero, and then parsing input \textit{OCLExpression} as a set of sub OCL expressions \textit{sub-formulas} and a set of logic connectors \textit{connectors}, and using \textit{lastn} to record the last element of \textit{sub-formulas}, and iterating each formula \textit{s} of the \textit{sub-formulas} set. In each iteration, algorithm initializes \textit{num} to zero, which indicates whether the appropriate rule is found. According to the value of the tag, different rules are used to check whether the formula \textit{s} is matched, then assign the match rule number to \textit{num} if any rule matched the formula \textit{s}. The formula \textit{s} will be mapped to primitive operation according to the rule. The transformation result is saved in \textit{r}. Note that the OCL expression of pre-condition is mapped into primitive operations connected with Java logic expressions. Therefore, the algorithm appends the transformation result \textit{r} with the connector \textit{connectors[i]} if the tag is \textit{pre-condition} and the index \textit{i} does not point to the last element. Otherwise, the transformation result \textit{r} will be appended to the final result \textit{rs} with a broken line. 

\begin{algorithm}
\SetKwData{SSD}{ssd}
\SetKwData{UC}{uc}
\SetKwInOut{Input}{Input}\SetKwInOut{Output}{Output}
\Input{OCLExpression, Tag}
\Output{Primitive Operations}

\Begin{ 

    \textit{rs} $\leftarrow$ $\emptyset$\;
    \textit{i}  $\leftarrow$ 0\;
    \textit{sub-formulas} $\leftarrow$ \textit{parse(OCLExpression)}\;
    \textit{connectors} $\leftarrow$ \textit{parseConnector(OCLExpression)}\;
    \textit{lastn} $\leftarrow$ \textit{len(sub-formulas)} - 1\;

    \For{s $\in$ sub-formulas}{
        \textit{num} $\leftarrow$ 0\;
        \Switch{Tag}{
            \Case{definition}{
                \textit{num} $\leftarrow$ \textit{matchRule1to7(s)}\;
            }  
            \Case{pre-condition}{
                \textit{num} $\leftarrow$ \textit{matchRule8to15(s)}\;
            } 
            \Case{post-condition}{
                \textit{num} $\leftarrow$ \textit{matchRule16to26(s)}\;
            } 
        }    
        \eIf{num != 0}{
            r $\leftarrow$ \textit{transform(s, num, Tag)}\;
            \eIf{tag == "pre-condition" and i != lastn}{
                \textit{rs.append(r, connectors[i])}\; 
            }{
                \textit{rs.append(r, "linebreaks")}\;
            }
        }{
            \textit{rs.append("transformation error for sub-formula:", s)}\; 
        }
        i++\;
    }    
    \Return \textit{rs}\;
}  
  \caption{Transformation Algorithm}
 \label{MappingAlgorithm}
\end{algorithm}

If no rule is matched to the formula \textit{s}, an error message will be appended to the final result \textit{rs} with the formula \textit{s}. The nature of requirements model implies that, in general, a requirements model may contain non-executable elements, because it focuses on what the system should do rather than how it does it. This error message will help to validate the requirements when generating the prototype. In the end of the iteration, \textit{i} is increased to point to the next sub-formula. After the iteration, the final result \textit{rs} will be returned. 

%
%
\subsection{Transformation Example}

We have already introduced the rules for transformation the contracts of system opeartions to the primitive operations. This subsection will present an example of \textit{enterItem()} of use case \textit{processSale} to explain how it work. Note that if a transformation rule is used more than once, we only explain the transformation at the first time. The signature of contract \textit{enterItem} is compiled to the operation signature in Java. 

In the definition section, the rule $R_3$ maps the OCL expression to the primitive operation \textit{findObject()} to find the object \textit{item} with the input barcode. In the pre-condition section, the rules $R_8$ and $R_{12}$ map the OCL contract to check whether the objects \textit{currentSale} and \textit{item} are undefined, the value of the attribute \textit{IsComplete} is false or not, and the attribute \textit{StockNumber} of object \textit{item} is great than zero. 

In the post-condition section, 1) the rule $R_{16}$ maps \textit{let..in} and \textit{oclIsnew()} to the primitive operation \textit{createObject()} to create object \textit{sli} of the class \textit{SaleLineItem} for the scanned product. 2) The rule $R_{19}$ maps \textit{includes} OCL contract to the primitive operation \textit{addOnetoManyAssociation()} by adding the link of the new created object \textit{sli} to the current sale object \textit{currentSale}. 3) The rule $R_{21}$ maps \textit{sli.BelongedItem = item} to the primitive operation \textit{addOnetoOneAssociation()} by linking the association \textit{BelongedItem} of object \textit{sli} to the object \textit{item}. 4) The rule $R_{23}$ maps \textit{sli.Subamount = item.Price*quantity} to the primitive operations \textit{setAtrribute()} and \textit{getAtrribute()} by setting the attribute \textit{Subamount} of object \textit{sli} to the sub amount value of \textit{item.getPrice() * quantity}. 5) The rule $R_{17}$ maps the \textit{includes} expression to the primitive operation \textit{addObject()} by adding the reference \textit{sli} to the records of class \textit{SalesLineItem}. 6) The rule $R_{25}$ maps the result expression to the Java return expression. All rules will be repeatedly applied for transforming until all the OCL contract are mapped to the primitive operations with Java code.
This section presented the transformation rules and algorithms for system operation decomposition. Next section will introduce how to generate prototype from the requirements model.





\section{Prototype}
\begin{figure}[!htb]
  \centering
  \includegraphics[width=0.48\textwidth]{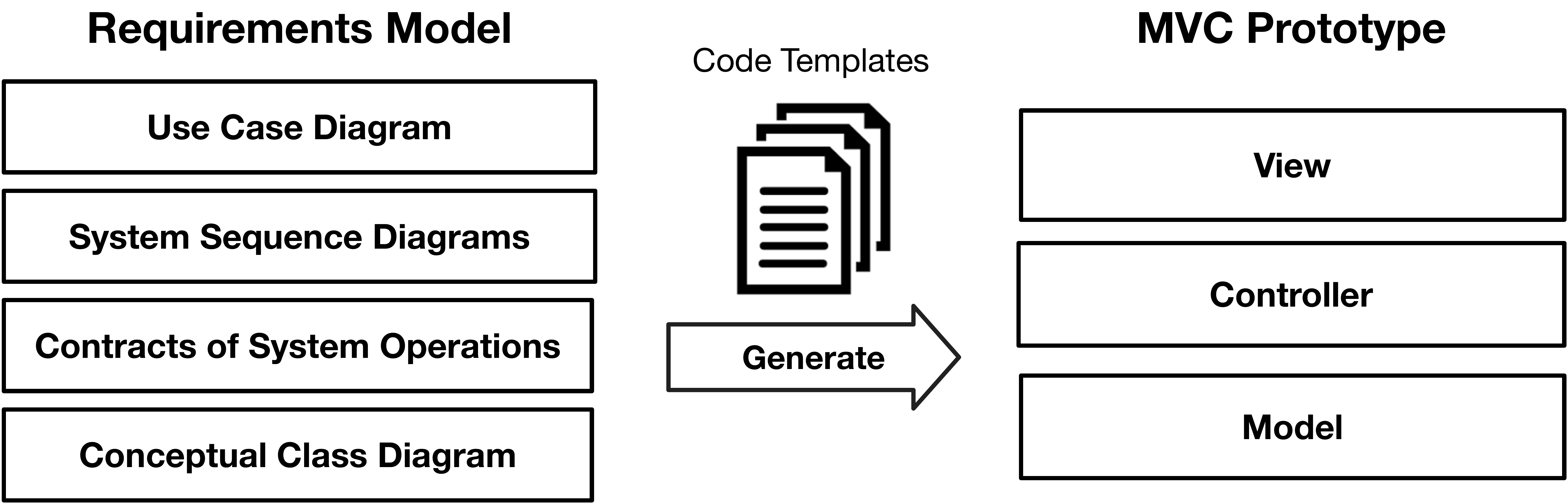}
  \caption{MVC Prototype Generation from Requirements Model}
  \label{prototype}
\end{figure}

Graphical User Interface (GUI) is one of the important parts of a prototype for customers and domain experts to validate their requirements. Model-View-Controller (MVC) \cite{Krasner:1988:CUM:50757.50759} is one of the most widely used GUI architecture patterns in decades. In this paper, we apply the MVC architecture pattern to prototype, which decouples views and models to make a clear division between models and GUI elements. The prototype contains three modules: view, controller, and model as shown in Figure \ref{prototype}.

%
%

\begin{figure*}[!htb]
  \centering
  \includegraphics[width=0.95\textwidth]{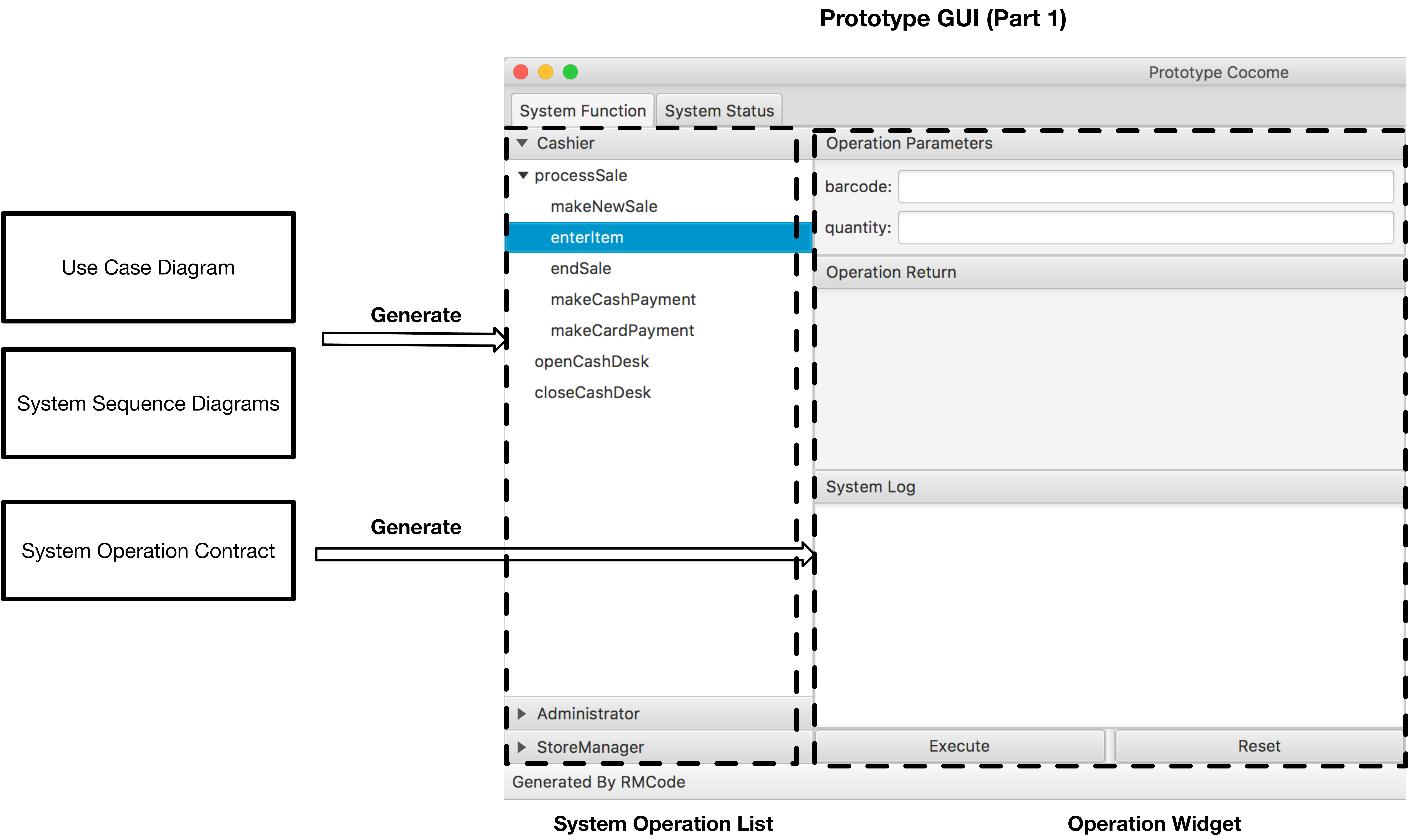}
  \caption{Prototype GUI for System Function}\label{prototypesf}
\end{figure*}

\begin{figure*}[!htb]
  \centering
  \includegraphics[width=0.95\textwidth]{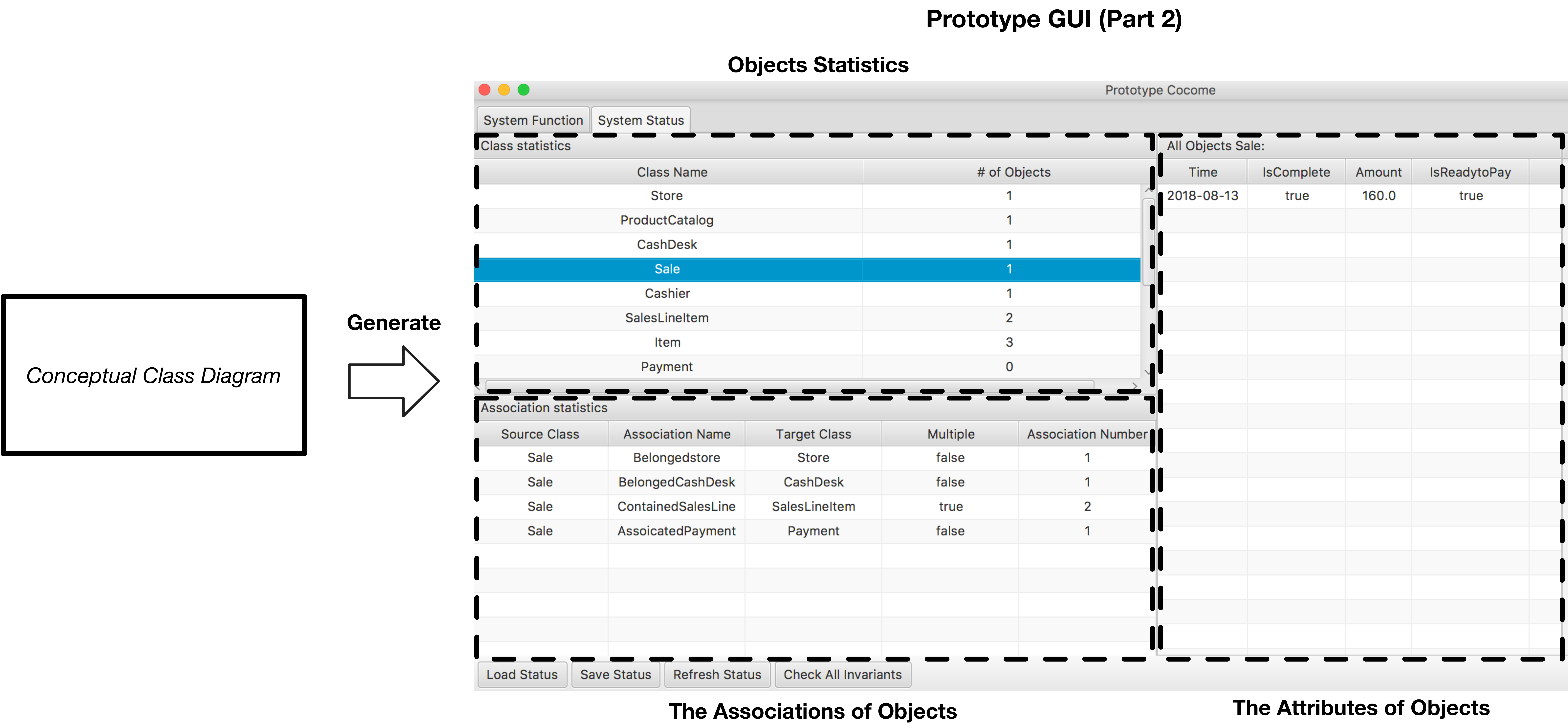}
  \caption{Prototype GUI for System State}\label{prototypess}
\end{figure*}

%
\vspace{.1cm}
\noindent \textbf{View:} The view module of the prototype contains UI widgets for end-users to validate system operations and observe corresponding changes in the system state. It contains two UI panels: system functionality and system state. As an example, the system functionality panel of CoCoME is shown in Figure \ref{prototypesf}. It includes the widgets of \textit{system operation lists} in the left side, and the \textit{operation widgets} for passing the parameters to system operations and returning the result in the right side. \textit{System operation lists} are generated from the use case diagram and the system sequence diagram. Moreover, the \textit{operation widgets} for inputting the parameters of system operations are generated from the signatures of the contracts of system operations. 

The system state panel is used to observe the state of objects of a system before and after executions of system operations. Figure \ref{prototypess} shows the state panel of CoCoME. The system state panel is generated from the conceptual class diagram. The left top side widget of \textit{objects statistics} presents the name and the number of the objects for each class in the current state. The left bottom side widget of \textit{the association of objects} shows the state of associations, which includes the source object, the target object, the name of the association, the number of the associated objects, and whether this association is a one-to-one or one-to-many relationship. When users click a class entry on the left side, the state of the corresponding attribute will display on the right panel widget of \textit{the attributes of objects}.

%
\vspace{.1cm}
\noindent \textbf{Controller:} The controller module links the view and model modules, which makes UI events to trigger system operations. The controller listens to the events from UI widgets. When a specific event is captured, the controller retrieves the input parameters from the UI widget, and then deliver the parameters to the corresponding system operation in the model module of the prototype. Finally, the controller will update the UI widgets with the return result. For example, when a cashier clicks the \textit{execute} button on the operation widget of Figure \ref{prototypesf}, the controller will capture a button clicked event generated by the UI widget, and then the controller will get the input parameters and then invoke the \textit{enterItem()} operation. Finally, the controller will update the UI widgets with the return result, and then the cashier will see the result from \textit{operation return} panel of Figure \ref{prototypesf}.

\vspace{.1cm}
\noindent\textbf{Model:} The model module is the core of the MVC architecture pattern. This module contains 1) the classes encapsulating system operations, which can be invoked by the controller with the parameters from UI widgets, and return the result to the controller for displaying on the view widgets, and 2) the classes generated from the conceptual class model. Note that we name the classes generated from the conceptual class model as \textit{entity classes}, and the others as \textit{fabricated classes}. We will show how to generate fabricated and entity classes in the remaining of this section.

\subsection{Fabricated Class Generation}
\begin{figure}[!htb]
  \centering
  \includegraphics[width=0.45\textwidth]{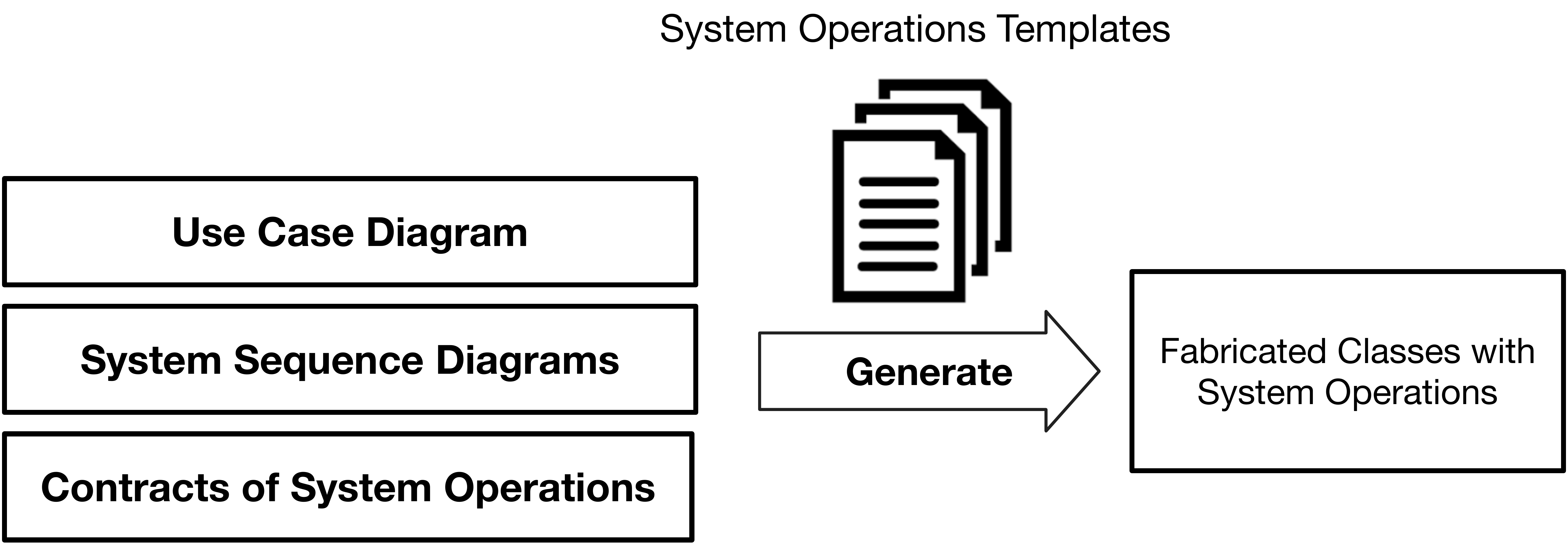}
  \caption{Code Generation for System Operation}\label{codesystemoperation}
\end{figure}
Fabricated classes encapsulating system operations are generated from a use case diagram, system sequence diagrams, the contracts of system operations through system operation templates in Figure \ref{codesystemoperation}. We already discussed the system operation decomposition in the previous section. To implement the decomposed system operation, we not only need to transform the contracts of system operations into primitive operations through the transformation algorithm and rules but also need to orchestrate them to be a validated system operation like the decomposition example of \textit{enterItem()} in Figure \ref{mapEnterItem}. Moreover, the implemented system operations must be encapsulated into classes. In this subsection, we introduce system operation templates to generate the implementations of the system operations from contracts, and how to encapsulate them into classes.


\subsubsection{System Operation Template}
Xtend \cite{bettini2016implementing} is a flexible and expressive template language. In this paper, we adopt Xtend to define the template. The template used to generate the implementation of system operations from the contract \textit{c} is shown as follows: 

\begin{lstlisting}[language=mytemplate]
/* operation signature */
public «c.opSign.returnType» «c.opSign.name»(
«FOR para : c.opSign.parameters SEPARATOR ','»
    «para.type» «para.name»
«ENDFOR») throws PreconditionException {

/* contract definition */
«IF c.definition != null»
    «c.definition.mapping»
«ENDIF»

/* check precondition */
if («c.precondition.mapping») { 

    /* contract post-condition*/
    «c.postcondition.mapping»

/* result return */
«IF c.opSign.returnType != null»
    return «returnName»; 
«ENDIF»
else {
    throw new PreconditionException();				
}}
\end{lstlisting}

Note that the template will be directly as the output result from the prototype generation, except interpolated expressions inside guillemets \textit{\flqq \frqq}. The expressions in \textit{\flqq \frqq} will be dynamically interpreted in terms of requirements model. Accordingly, the expressions between \textit{\flqq IF\frqq} and \textit{\flqq ENDIF\frqq} will be interpolated when the condition is evaluated as \textit{true}. The expressions between \textit{\flqq FOR\frqq} and \textit{\flqq ENDFOR\frqq} will be repeatedly interpolated through all the elements of the list. For example, if $c$ represents the contract of \textit{enterItem()}, \textit{\flqq c.opSign.name\frqq} will be interpreted as the name of the system operation \textit{enterItem}.

The above template helps to implement the contract signature through generating the Java operation with keyword \textit{public}, the name of system operation \textit{\flqq c.opSign.name\frqq}, input variables \textit{\flqq para.name\frqq} with type \textit{\flqq para.type\frqq}, and return type \textit{\flqq c.opSign.returnType\frqq}. Then the template helps to generate the code \textit{\flqq c.definition.compile\frqq} for the definition section of the contract, if any. Then, the template generates the Java \textit{if-else} control flow to check the pre-condition of system operation \textit{\flqq c.precondition.compile\frqq}, and executes the logic code \textit{\flqq c.postcondition.compile\frqq} to make the system satisfying the post-condition. In detail, 1) if the evaluation result of pre-condition is true, the code of the system operation will be executed, and then return the result by Java code \textit{return \flqq returnName\frqq}, if any. 2) If the evaluation result of pre-condition is not true, an exception \textit{PreconditionIsNotSatified} will be directly emitted without executing any code of the system operation. However, this template only helps to generate the skeleton code of system operation. The \textit{\flqq c.definition.mapping\frqq}, \textit{\flqq c.precondition.mapping\frqq} and \textit{\flqq c.postcondition.mapping\frqq} are interpreted by transformation rules and algorithm in section 3. 

\subsubsection{System Operation Encapsulation}
%
%
System operations are captured in system sequence diagrams of use cases. If we follow the suggestion of the expert pattern in GRASP, system operations should be encapsulated to the classes (experts) in the conceptual class model. For example, the class \textit{Item} contains attributes of \textit{barcode} and \textit{price}, \textit{enterItem(barcode, quantity)} should be assigned to the class \textit{Item}. However, entity classes have already held primitive operations for manipulating the attributes and associations. We expect to separate concerns into a different abstract level of class to achieve high cohesion and low coupling. 
\begin{figure}[!htb]
  \centering
  \includegraphics[width=0.35\textwidth]{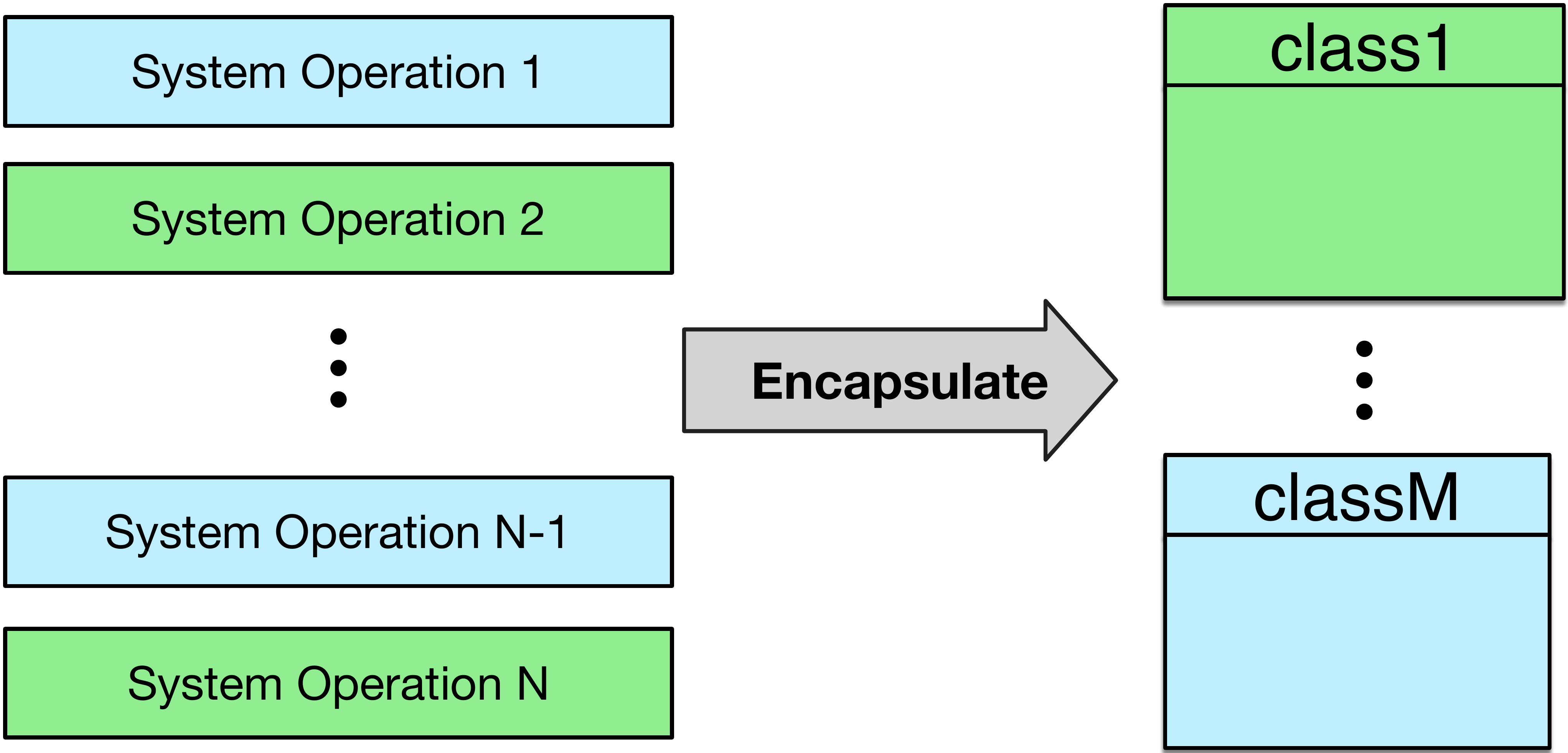}
  \caption{System Operation Encapsulation}\label{assignment}
\end{figure}

The mediator pattern of GoF suggests an object that encapsulates a set of operations to promote loose coupling. That describes the situation the same as a use case, which includes high cohesion system operations related to one scenario of interactions between the actor and the system. Therefore, we generate a pure fabrication class for each use case and then encapsulate the system operations of the use case to the fabrication class to achieve high cohesion. 

\begin{algorithm}
\SetKwData{SSD}{ssd}
\SetKwData{UC}{uc}
\SetKwInOut{Input}{Input}\SetKwInOut{Output}{Output}
\Input{\textit{ucd} - Use Case Diagram, \\
\textit{ssds} - System Sequence Diagrams, \\
\textit{contracts} - Contracts}
\Output{\textit{ucClasses} - fabrication classes}
\tcc{Initialize ucClasses as empty set} 
\textit{ucClasses} $\leftarrow$ $\emptyset$ \;
\For{uc $\in$ ucd}{
    \tcc{generate fabrication uc class} 
    \textit{ucClass} $\leftarrow$ \textit{generateClassSkeleton(uc, classes)}\;
    \tcc{find uc related system sequence diagram} 
    \textit{ssd} $\leftarrow$ \textit{findSSD(uc, ssds)}\;
	\For{op $\in$ ssd}{
	     \tcc{find operation signature}
	     \textit{opSign} $\leftarrow$ \textit{findSignature(op, contracts)}\;
	     \tcc{encapsulate operation to ucClass}
	     encapsulate \textit{opSign} to class \textit{ucClass}\;
    }
    add \textit{ucClass} to \textit{ucClasses}\; 
  }
  
  \caption{Generation Algorithm for Fabricated Class and System Operations Encapsulation}
 \label{collborationsAssignment}
\end{algorithm}

Algorithm \ref{collborationsAssignment} generates the fabrication classes and encapsulates system operations into them.  However, if a target problem contains many use cases, and most use cases only include one system operation, Algorithm \ref{collborationsAssignment} will generate many fabrication classes with just one operation. That breaks the design principle of high cohesion. The appropriate way is to define one fabrication class charging for several responsibilities to promote high cohesion. In the extreme cases, you can even specify only one pure fabrication class for all use cases. To deal with those situations, our CASE tool provides a mechanism that allows product managers and domain experts to decide how to define those fabrication classes for the use cases, and then automatically generates the implementation of those pure fabricated classes.

\begin{figure}[!htb]
  \centering
  \includegraphics[width=0.45\textwidth]{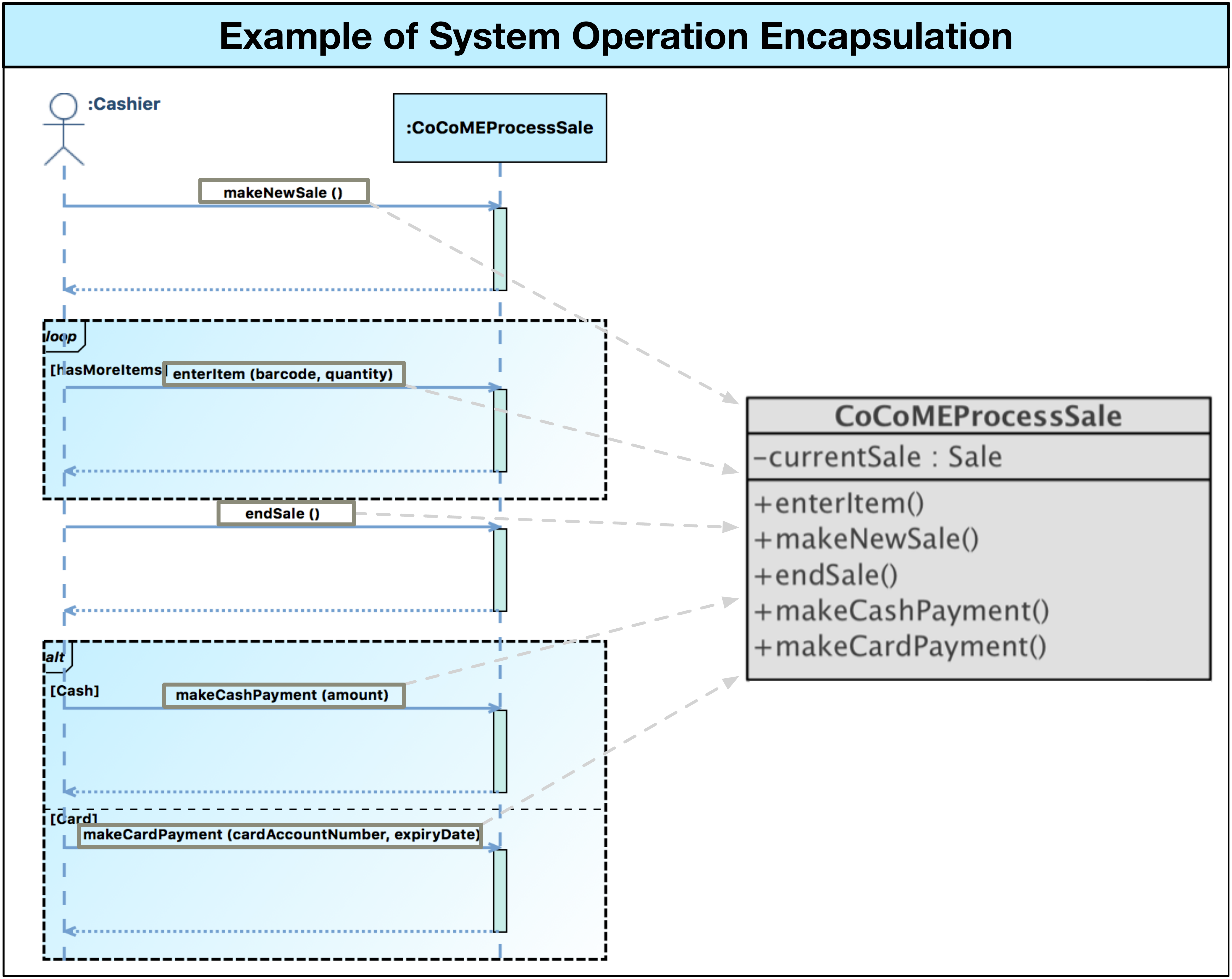}
  \caption{System Operation Encapsulation of Use Case ProcessSale}\label{assignmentexample}
\end{figure}

For example, the \textit{currentSale} object of class \textit{Sale} created by the system operation \textit{makeNewSale()} is reused in the system operations \textit{enterItem()} and \textit{endSale()} of the use case \textit{processSale}. Algorithm \ref{collborationsAssignment} generates a pure fabricated class \textit{CoCoMEProcessSale} to encapsulate those system operations and the shared object \textit{currentSale}. In detail, \textit{CoCoMEProcessSale} is the pure fabrication class for encapsulating system operations \textit{enterItem()}, \textit{makeNewSale()}, \textit{endSale()}, \textit{makeCashPayment()}, \textit{makeCardPayment()} and the object reference to \textit{currentSale} in Figure \ref{assignmentexample}.

This subsection illustrates how to generate fabricated classes and encapsulate the system operations. The next two subsections introduce how to generate entity classes from the conceptual class diagram as well as primitive operations.

\subsection{Entity Class Generation}
\begin{figure}[!htb]
  \centering
  \includegraphics[width=0.45\textwidth]{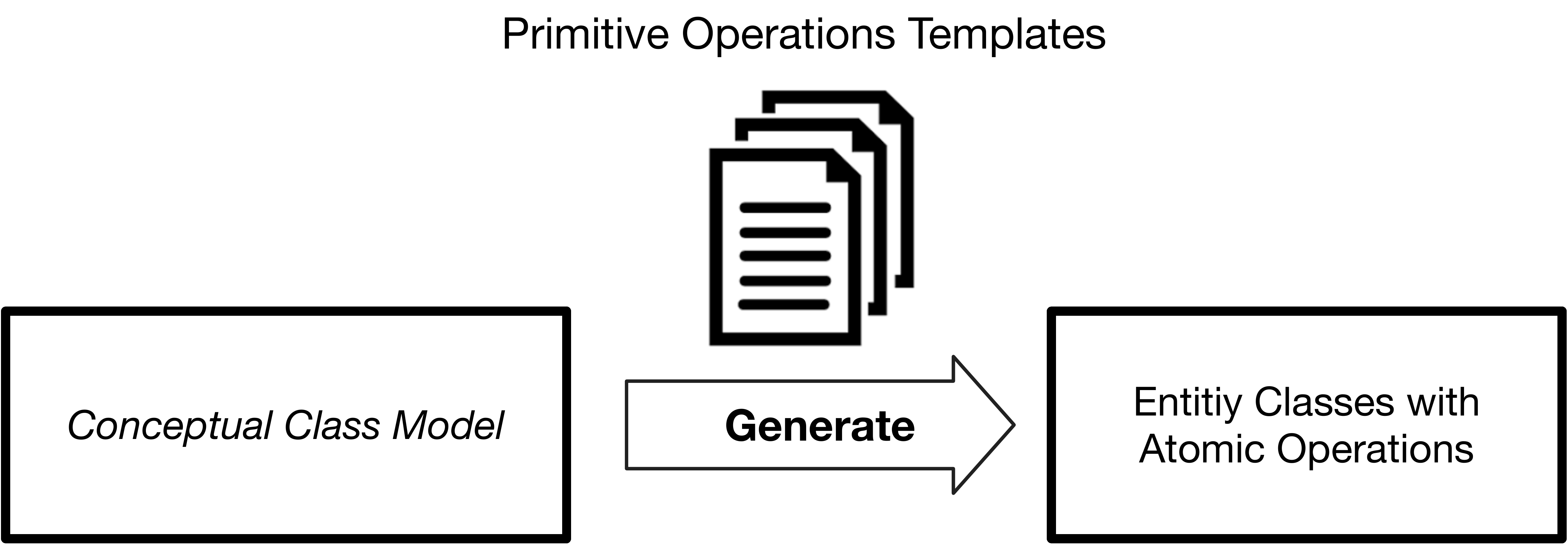}
  \caption{Code Generation for Primitive Operations}\label{codegenerationatomicoperation}
\end{figure}

%
%
Entity classes generation is a necessary procedure to achieve the auto-prototyping from requirements model, which is shown in Figure \ref{codegenerationatomicoperation}. It not only needs to generate the attributes and associations of the entity class but also requires to generate the implementation of primitive operations for setting and getting the attributes, finding linked objects, and adding and removing the links. We will show the details in the subsection.

\subsubsection{Generation Algorithm for Entity Classes}
The information expert pattern of GRASP suggests assigning responsibilities to the information expert, and the expert class knows the information necessary to fulfill the responsibilities. Therefore, entity classes encapsulate primitive operations about getting and setting the values of the their attributes, forming and breaking the their links. For example, entity class \textit{Item} has the \textit{Price} attribute, therefore, primitive operation \textit{getPrice()} is encapsulated in class \textit{Item}. Entity class \textit{SaleLineItem} has the \textit{Quantity} attribute and the association \textit{BelongedSale}, therefore, it encapsulates primitive operation \textit{updateAttributeQuantity()} and \textit{addAssociationTosale()}. The algorithm that generates the entity classes from a conceptual class diagram is shown in Algorithm \ref{atomicAttriandAsso}.

\begin{algorithm}
\SetKwData{Left}{left}\SetKwData{This}{this}\SetKwData{Up}{up}
\SetKwFunction{Union}{Union}\SetKwFunction{FindCompress}{FindCompress}
\SetKwInOut{Input}{Input}\SetKwInOut{Output}{Output}

\Input{$ccd$ - Conceptual Class Diagram \\
$t_{ec}$ - Entity Class Template \\
$t_{ao}$ - Primitive Operation Templates}
\Output{Entity Classes}
    \For{entity $\in$ $ccd$}{
      generate entity class skeleton by $t_{ec}$\;
	  \For{attribute $\in$ entity}{
	     genereate \textit{getAttribute()} by $t_{ao}$\;
	     genereate \textit{setAttribute()} by $t_{ao}$\;
      }
      \For{association $\in$ entity}{
         \uIf{Is-Multiple(association) == true}{
            genereate \textit{findLinkedObjects($t_{ao}$)}\;
            genereate \textit{addLinkOnetoMany($t_{ao}$)}\;
            genereate \textit{removeLinkOnetoMany($t_{ao}$)}\;
         }
         \Else{
            genereate \textit{findLinkedObject($t_{ao}$)}\;
            genereate \textit{addLinkOnetoOne($t_{ao}$)}\;
            genereate \textit{removeLinkOnetoOne($t_{ao}$)}\;
         }
      }
    }
  \caption{Generation Algorithm for Entity Classes}
 \label{atomicAttriandAsso}
\end{algorithm}

This algorithm takes a conceptual class model as input and generates entity classes with primitive operations. For each class in the conceptual class diagram, entity class skeleton is generated through the entity class template. Then, primitive operations about getting and setting each attribute of the entity class are generated. For each association of the class, a) if the association is a one-to-many relationship, primitive operations about finding linked objects, adding and removing link of one-to-many association are generated, b) if the association is a one-to-one relationship, primitive operations about finding the linked object, adding and removing link of one-to-one association are generated. Note that above entity skeleton and primitive operations are generated through the code templates, and we will introduce those templates in the remaining of this subsection.

%
%
\subsubsection{Entity Class Template} 
The generation skeleton of entity classes is straightforward. Many UML tools support this feature. The template to generate an entity class \textit{c} is: 
\begin{lstlisting}[language=mytemplate]
/* Class Skeleton */
public class «c.name»

/* Class Inheritance */
«IF c.superClass != null» 
    extends «c.superClass.Name» 
«ENDIF» {
   
/* Attributes */
«FOR attribute : c.attributes»
    private «attribute.type» «attribute.name»;
«ENDFOR»
   
/* Associations */
«FOR assoc : c.associations»
    private 
    «IF assoc.isIsmultiple»
        List<«assoc.class»> «assoc.name» = 
            new LinkedList<«assoc.class»>();
    «ELSE»
        «assoc.class.name» «assoc.name»;
    «ENDIF»
«ENDFOR»

/* primitive operations templates */
}
\end{lstlisting}
The template defines a public class with the keyword \textit{public} and the name \textit{c.name} inside of «». If the class has a super-class, the text of \textit{c.superClass.name} will be replaced by the name of the super-class along with the keyword \textit{extends}. The attribute is declared by the keyword \textit{private} with the type \textit{attribute.type} and the name \textit{attribute.name}. The association is also declared by the keyword \textit{private} with a) a typed list \textit{List<assoc.class>} and association name \textit{assoc.name}, if the association is a one-to-many relation, b) an attribute with the name of associated class \textit{assoc.class.name} and the type of associated class \textit{assoc.class}, if the association is a one-to-one relation.

%
%


\subsubsection{Primitive Operation Templates} 
The templates for primitive operations \textit{getAttribute()} and \textit{setAttribute()} are: 
\begin{lstlisting}[language=mytemplate]
//Getting Attribute
public «attribute.type» get«attribute.name»() {
    return «attribute.name»;
}	

//Setting Attribute
public void set«attribute.name»
            («attribute.type» «attribute.name») {
    this.«attribute.name» = «attribute.name»;
}
\end{lstlisting}
Getting an attribute can be defined as the public operation with the name \textit{get\flqq attribute.name\frqq} and the type \textit{\flqq attribute.type\frqq} of attribute, which is fulfilled by directly returning the attribute by using the keyword \textit{return}. The setting attribute can be fulfilled by setting the value of attribute \textit{this.\flqq attribute.name\frqq} with an input of typed variables.

%
%
The primitive operation templates for finding linked objects, linking a new object, and removing an linked object are: 
\vspace{.1cm}
\begin{lstlisting}[language=mytemplate]
//findLinkedObjects()
public List<«assoc.class»> get«assoc.name»() {
    return «assoc.name»;
}	

//addLinkOnetoMany()
public void add«assoc.name»(«assoc.class» ob) {
    this.«assoc.name».add(ob);
}

//removeLinkOnetoMany()
public void remove«assoc.name»(«assoc.class» ob) {
    this.«assoc.name».remove(ob);
}
\end{lstlisting}
We use a reference list to store object references for the one-to-many association. Therefore, finding objects through link can be directly implemented by returning the reference list of the association. Forming a link can be fulfilled by invoking the adding operation of the list as well as the deleting operation for breaking a link.
 
\begin{lstlisting}[language=mytemplate]

//findLinkedObject
public «assoc.class» get«assoc.name»() {
    return «assoc.name»;
}	

//addLinkOnetoOne() removeLinkOnetoOne()
public void set«assoc.name»(«assoc.class» ob) {
    this.«assoc.name» = ob;
}			
\end{lstlisting}
The one-to-one association is implemented as an attribute with the type of associated class. Thus, the implementation of finding the object through a link of one-to-one association is the same as getting the value of an attribute. The implementation of adding and removing a link of an one-to-one association is the same as setting primitive operation of an attribute. For example, when removing a link of a one-to-one association, we can just set \textit{null} as the value of the link.

%
%
\subsection{EntityManager Generation} 

In object-oriented system, the object can be eventually found through the links of the objects. However, we need to locate the entrance object then to start the finding procedures. The primitive operation \textit{findObject()} is used to find the first entrance object, then other objects can be reached through the primitive operation \textit{findLinkedObject()} of the links of objects. To implement \textit{findObject()}, a pure fabrication class named as \textit{EntityManager} is required, which contains all the references to the objects of entity classes as well as the primitive operations \textit{findObject()} and \textit{findLinkedObject()}. Moreover, the creator pattern of GRASP suggests assigning the creating responsibility to the class recording the instances of the created objects. \textit{EntityManager} records all instances of entity classes. Therefore, the primitive operations \textit{createObject()} and \textit{releaseObject()} are assigned to \textit{EntityManager}. In addition, \textit{EntityManager} is required a global accessing point. According to the singleton pattern of GoF, we build \textit{EntityManager} as a singleton class, which has the only one instance, and provide a global accessing point. In short, the generation algorithm of \textit{EntityManger} is shown in Algorithm \ref{entitymanager}.



\begin{algorithm}
\SetKwData{Left}{left}\SetKwData{This}{this}\SetKwData{Up}{up}
\SetKwInOut{Input}{Input}\SetKwInOut{Output}{Output}
\Input{$ccd$ - Conceptual Class Diagram \\
$t_{em}$ - EntityManager Template \\
$t_{o}$ - Primitive Operation Templates for Object}
\Output{EntityManager Class}
\Begin{
  \tcc{Generate EntityManager Skeleton}
  generate \textit{EntityManager} skeleton by $ccd$, $t_{em}$\;
  \tcc{Generation Primitive Operations}
  generate \textit{findObject() by $t_{o}$}\;
  generate \textit{findObjects()} by $t_{o}$\;
  generate \textit{createObject()} by $t_{o}$\;
  generate \textit{addObject()} by $t_{o}$\;
  generate \textit{releaseObject()} by $t_{o}$\;
  }
  \caption{Generation Algorithm for EntityManager}
 \label{entitymanager}
\end{algorithm}

This algorithm takes a conceptual class model, \textit{EntityManager} template, the templates of object-related primitive operation as input, generates \textit{EntityManager} classes. We generate \textit{EntityManager} skeleton  through \textit{EntityManager} template, and the implementation of primitive operations \textit{findObject()}, \textit{findObjects()}, \textit{createObject()}, \textit{addObject()} and \textit{releaseObject()} inside of \textit{EntityManager}. Note that \textit{EntityManager} records all the instances of entity classes, and \textit{addObject()} is used for adding the object into the records. We will show those templates in the remaining of this subsection.
\subsubsection{EntityManager Template} 
The template for generating \textit{EntityManager} skeleton is shown as follows: 

\begin{lstlisting}[language=mytemplate]
/* EntityManager Template */
public class EntityManager {

    /* HashMap Object Records*/
    private static Map<String, List> AllInstance = new HashMap<String, List>();

    /* create object reference list */
    «FOR c : classes»
    private static List<«c.name»> «c.name»Instances = 
        new LinkedList<«c.name»>();
    «ENDFOR»

    /* Put object reference list into Map */
    static {
    «FOR c : classes»
        AllInstance.put("«c.name»", «c.name»Instances);
    «ENDFOR»
    }
    
    /* Get all objects of the class   */
    public static List getAllInstancesOf
        (String ClassName) {
        return AllInstance.get(ClassName);
    }	
}
\end{lstlisting}
We use Java \textit{HashMap} named \textit{AllInstances} to record all the object reference lists. Each object reference list is implemented as a \textit{LinkedList} of Java with name \textit{\flqq c.name\frqq Instances}. \textit{\flqq c.name\frqq} will be replaced by the name of classes in the conceptual class model. All references list of object are added to this \textit{HashMap} \textit{AllInstance}. After that, we can find the specific object reference list directly by the name of entity class.

\subsubsection{Primitive Operation Templates for Finding Objects} 
By using \textit{EntityManager}, we can do fine-grained search \textit{findObject()} and \textit{findObjects()} with a query condition \textit{precondition(o)} on the reference list. The template of finding object is:
\begin{lstlisting}[language=mytemplate]
/* find object template */
«cName» target = null; //initialize target object
for («cName» o:
    EntityManager.getAllInstancesOf(«cName»)) {
    //finding the object satisfies the condition 
    if («precondtion(o)») {
        target = o;
        return target;
    }
}
\end{lstlisting}
The template initializes the object reference \textit{target} as \textit{null}, and then iterates the object list by \textit{getAllInstancesOf} to find the object \textit{o} satisfying the search condition \textit{precondition(o)} (refer the details of  precondition transformation to section 3.2.2), Finally, the template assigns the finding object \textit{o} to object \textit{target}, and returns the found object. The template \textit{findObjects} is similar to \textit{findObject}, which is:
\begin{lstlisting}[language=mytemplate]
/* find objects template */
List<«c.name»> targets = = new LinkedList<>(); //initialize target object lists
for («c.name» o:
    EntityManager.getAllInstancesOf(«c.name»)) {
    //finding the object satisfies the condition 
    if («precondtion(o)») {
        targets.add(o);
    }
}
return targets;
\end{lstlisting}
The differences is the \textit{findObjects()} template initializes the \textit{targets} as a linked list, 
adds the object \textit{o} to the list \textit{targets} when the object \textit{o} satisfies the condition, and finally, returns the target list \textit{targets}.

\subsubsection{Templates for Creating, Adding and Releasing Object} 
The factory method pattern of GoF suggests defining an interface for object creation, but let sub-classes decide which class to instantiate. Therefore, we impact all the create object responsibilities in the \textit{EntityManager}, and define a factory method that creates all objects of classes. This factory method can invoke a concrete creator that returns an instance of an entity class through the Java reflect mechanism. The primitive operation template for creating objects is: 
\begin{lstlisting}[language=mytemplate]
/* create object template */  
public static Object createObject(String cName) {
    Class c = Class.forName("EntityManager");
    Method m = c.getDeclaredMethod("create" + cName + "Object");
    return m.invoke(c);
}

«FOR c : classes»
public static «c.name» create«c.name»Object() {
    «c.name» o = new «c.name»();
    return o;
}
«ENDFOR»
\end{lstlisting}
The factory method \textit{createObject} provides a single-point to create object, it will invoke the concrete \textit{create«c.name»Object} method to create and return the object of class \textit{«c.name»}. We also use this pattern to build the \textit{addObject} and \textit{deleteObject} template. The template for generating primitive operation \textit{addObject()} is
\begin{lstlisting}[language=mytemplate]
/* add object template */  
public static Object addObject
    (String cName, Object ob) {

    Class c = Class.forName("EntityManager");
    Method m = c.getDeclaredMethod("add" + cName + "Object", Class.forName(cName));
    return  (boolean) m.invoke(c, ob);

}	

«FOR c : classes»
public static boolean add«c.name»Object(«c.name» o) {
    return «c.name»Instances.add(o);
}
«ENDFOR»
\end{lstlisting}
Object \textit{ob} and the name of class are passed to the factory method \textit{addObject}. Then \textit{addObject} uses Java reflect to invoke the concrete method \textit{add«c.name»Object} to add the object \textit{ob} to the object list \textit{«c.name»Instances}. We use the same pattern to build the template \textit{deleteObject}:  
\begin{lstlisting}[language=mytemplate]
/* release object template */
public static boolean deleteObject
    (String cName, Object ob) {

    Class c = Class.forName("EntityManager");
    Method m = c.getDeclaredMethod("delete" + cName + "Object", Class.forName(cName));
    return  (boolean) m.invoke(c, ob);
}

«FOR c : classes»
public static boolean delete«c.name»Object
    («c.name» o) {
    return «c.name»Instances.remove(o); 
}
«ENDFOR»
\end{lstlisting}

%
%
\section{Requirements Validation and Evolution}
\label{sec:requirementvalidation}
\begin{figure}[!htb]
  \centering
  \includegraphics[width=0.4\textwidth]{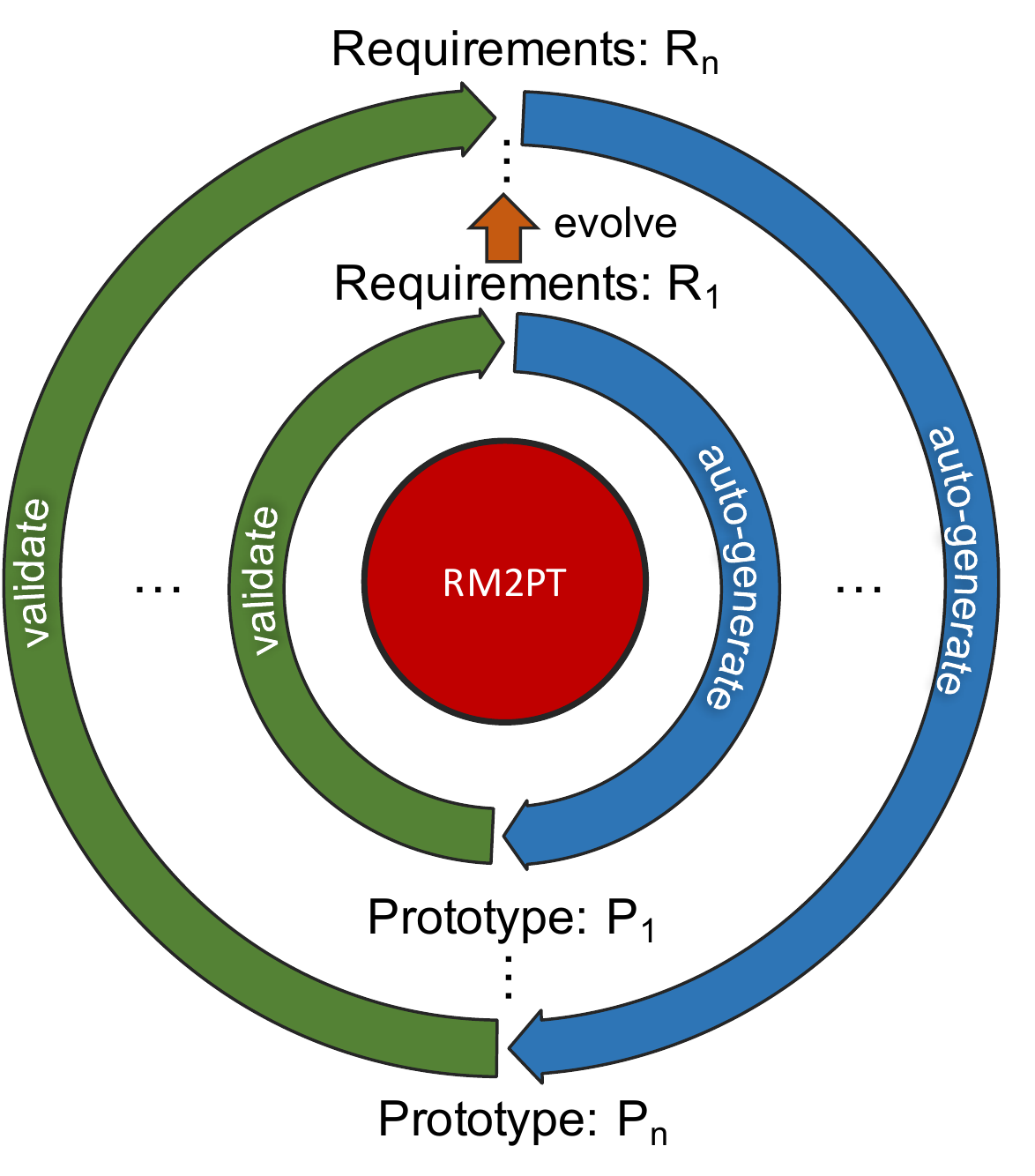}
  \caption{Requirement Validation and Evolution}\label{refinement}
\end{figure}

This section presents how to use the prototype to validate and evolve requirements. An overview is depicted in Figure \ref{refinement}.
 It starts with requirements $R_1$, from which a prototype $P_1$ is automatically generated using implementation CASE tool \textit{RM2PT} based on the techniques presented in Section 3 and 4. Then $P_1$ is used to validate the requirements $R_1$. If the faults of requirements or the missing requirements are found in $R_1$, we evolve the requirements $R_1$ to a requirements $R_2$ by fixing the errors and elicit new requirements. Then a prototype $P_2$ is generated from requirements $R_2$, we use prototype $P_2$ to validate requirements $R_2$. This process repeats until no more faults and missing requirements are found. In the practice, requirements are often a moving target, and thus requirements validation and evolution can be triggered in any stage of software engineering if needed.

\begin{figure*}[!htb]
  \centering
  \includegraphics[width=0.95\textwidth]{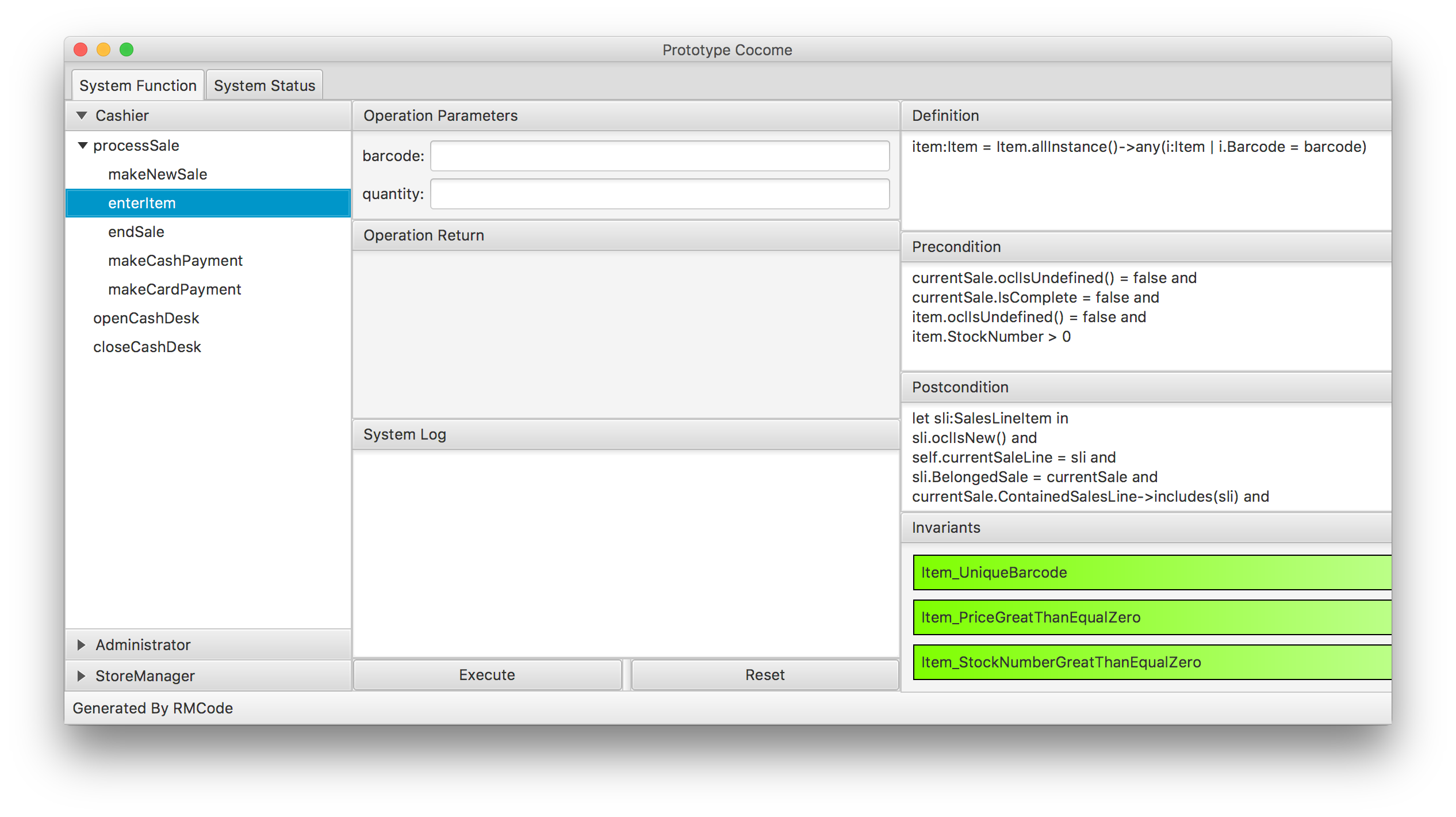}
  \caption{System Operation Validation}\label{systemfunctionality}
\end{figure*}

%
%
Customers and project managers can quickly validate functional requirements through the generated prototype without writing any code. A prototype provides two modules for requirements validation: 

\vspace{.1cm}
\noindent\textbf{$\bullet$ System Operation Execution with Pre-condition and Invariant Checking} \hspace{1mm} Prototype provides a window for the customers to test system operations. For example, CoCoMe prototype includes the widgets of system operation list on the left side, the system operation panel in the middle and the contracts panel on the right side in Figure \ref{systemfunctionality}. 
\begin{figure}[!htb]
  \centering
  \includegraphics[width=0.40\textwidth]{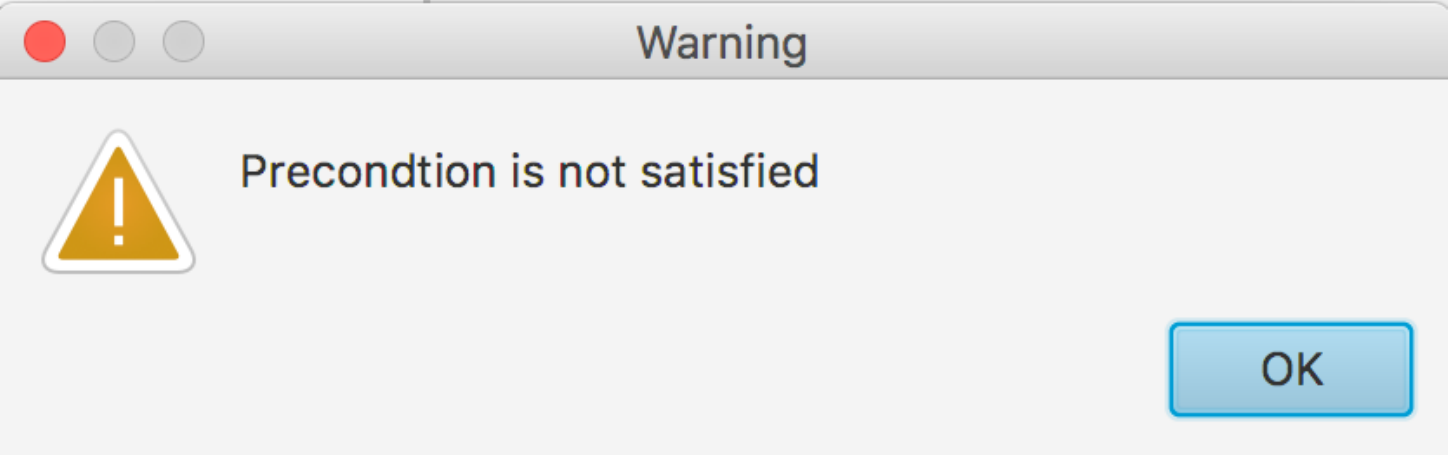}
  \caption{Pre-condition Checking}\label{precondition}
\end{figure}
 
\begin{figure}[!htb]
  \centering
  \includegraphics[width=0.5\textwidth]{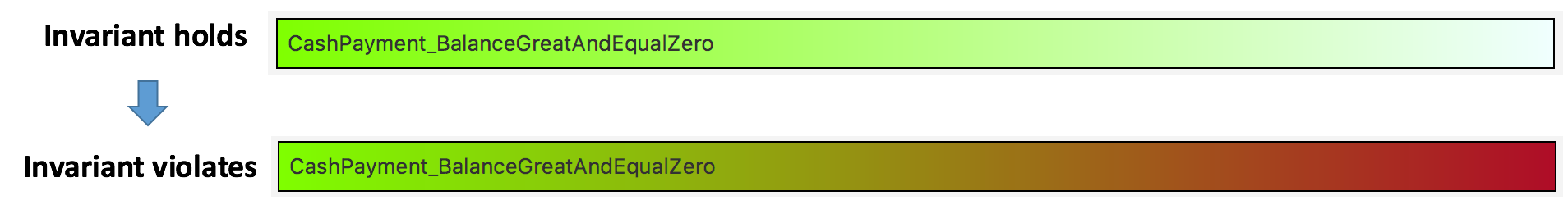}
  \caption{Invariant Checking}\label{invariant}
\end{figure}
Moreover, the prototype involves pre-conditions and invariants checking. When executing a system operation, a warning message is prompted if pre-condition is not satisfied in Figure \ref{precondition}. Moreover, if the execution of the system function make system state break the related invariants, the color of the invariant bar will become red from green in Figure \ref{invariant}. That indicates the errors in the requirements. Further inspections are required to locate the errors.

\begin{figure*}[!htb]
  \centering
  \includegraphics[width=0.95\textwidth]{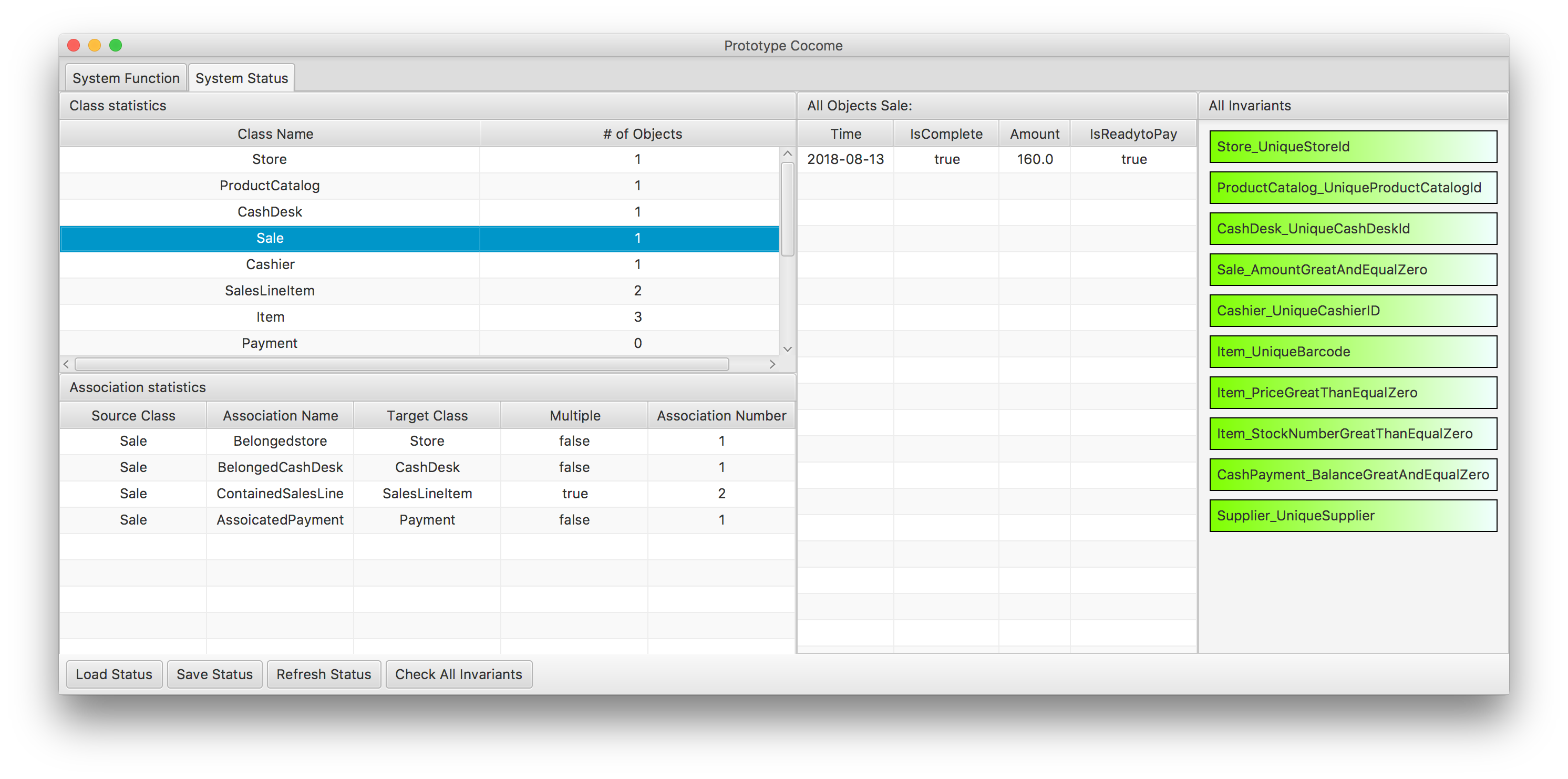}
  \caption{System State Observation}\label{systemstatus}
\end{figure*}

\vspace{.1cm}
\noindent\textbf{$\bullet$ State Observation} The location of the errors may be in the pre-condition, post-condition of the contract. Invariant checking can only indicate the errors. The state observation of the objects can help to locate the faults of the requirements. Figure \ref{systemstatus} shows the state observation panel of CoCoME prototype. The left top side presents the name and the number of the objects for each class in the current system. For example, the number of the objects of the classes \textit{Store}, \textit{CashDesk}, \textit{Sale} and \textit{Item} in the CoCoME system. The left bottom side shows the status association, which includes the source object, the target object, the name of the association, the number of the associated objects, and whether this association is a one-to-one or one-to-many relationship. For example, the Figure \ref{systemstatus} shows the state of association \textit{BelongedCashDesk} of the object \textit{Sale}. The middle side of the panel shows the status of the attributes of the objects. E.g, the status of the attributes \textit{Time}, \textit{IsComplete}, \textit{Amount} of the \textit{Sale} objects. When clicking a class entry on the left side, the state of the corresponding attributes and associations will display on the middle and left bottom side of the panel. Furthermore, all the invariants of the system are listed on the right side of the panel. When any invariant does not hold, it will become as red bar like in the system function panel. The remaining parts of this subsection will show how to use the generated prototype to validate the requirements.


%
%

%
%
\subsection{Start-up Objects}
When the prototype is opened for the first time, it does not contain any object. That means system operations except adding objects cannot be executed because the pre-conditions of functions are not satisfied. The start-up objects must be added into the prototype before requirements validation. 

\begin{figure}[!htb]
  \centering
  \includegraphics[width=0.5\textwidth]{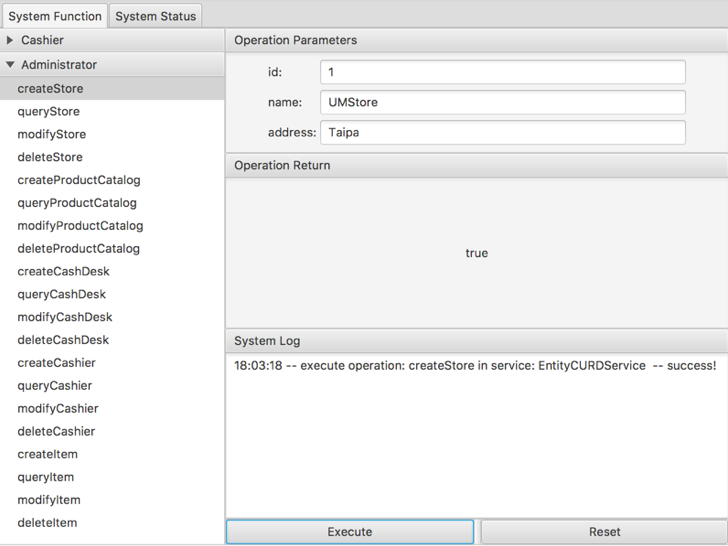}
  \caption{Start-up Objects}\label{startup}
\end{figure}

The prototype provides two options to load start-up objects. 1) Load objects from the checkpoint file. The users can click the \textit{Load State} button in the system state panel to load the checkpoint files into the system to restore the status of system saved before. 2) Manually create start-up object by using the Create, Read, Update, and Delete (CRUD) operations provided by prototypes. RM2PT provides mechanisms to automatically generate the contracts and implementations of CRUD operations by marking conceptual classes. The administrator can use this mechanism to add the start-up objects into the prototype. For example, administrators add the \textit{Store} object with the ID \textit{1}, name \textit{UMStore}, and address \textit{Taipa} into the system by \textit{createStore()} operation in Figure \ref{startup}.



%
%
\subsection{Pre-condition Validation}
The execution of system operation containing pre-condition errors may lead system to an un-expected state that violates system invariants. It contains two cases: parameter constraint missing and constraint error.

\vspace{.05cm}
\noindent \textbf{Parameter Constraint Missing:} Parameter constraint missing means that the constraints of parameters are missing in the pre-condition, the execution of the system operation may lead system to an un-expected state that violates system invariants. For example, the customer makes a cash payment \$20 for their \$40 products in CoCoMe. If the pre-condition does not check the tendered money is greater than or equal to the total price of products, the invariant of  \textit{CashPayment} that the balance must be greater than or equal to zero may be violated. The invariant bar of the prototype will become red as in Figure \ref{invariant}. Users should add the missed constraints into the pre-condition of \textit{cashPayment} to fix this error.

\vspace{.05cm}
\noindent\textbf{Constraint Error:} Constraint error means that the constraints are not correct in the pre-condition, the execution of the system operation may lead system to un-expected state that violates system invariants. For example, ATM has the invariant about the one-day maximum withdraw limitation (E.g., \$2000 pre-day). When a customer intends to withdraw \$500 from her account if she has already withdrawn \$1800 on the same day. This request must be rejected by the pre-condition of \textit{withdraw} operation. However, this request can be executed if the pre-condition is that the total withdrew money of one day is lesser than or equal to \$2000. In this case, the execution of \textit{withdraw} will make the system state violate the maximum limitation invariant due to the withdrew money \$2300 is higher than the maximum limitation \$2000. We need to fix this error by replacing the pre-condition to check the value of withdrawing must be lesser than or equal to the value of maximum limitation minus the total value of withdrew today ($withdrawing \le maxValue - withdrew$).


%
%
\subsection{Post-condition Validation}
Post-condition errors can also lead the system to an un-expected states that violate system invariants and the pre-condition of the next system operations under the same use case. For examples, 1) one invariant of CoCoME system shows the final price of the current sale process must be greater than zero. The operation \textit{endSale()} adds all the sub-amount of products together as the final price for the payment. If there is a typo when the plus sign '+' is typed as minus sign '-' in the post-condition, the final price will be less than zero. That will violate the price invariant after execution. 2) the post-condition of \textit{endSale()} should specify the flag \textit{IsReadytoPay} to \textit{true}. If missing this specification in the post-condition, the pre-condition of the system operation \textit{makePayment()} cannot be satisfied, and a warning message will prompt in Figure \ref{precondition}. The product managers can validate and fix those post-condition errors without too many efforts by checking the system state before and after the execution of system operations in the prototypes.

%
%

In short, the prototype can indicate the pre-condition and invariant violations in requirements models. Although locating and fixing the errors require carefully observing and analyzing the state of objects and the contracts, the generated prototype provides an intuition way to help product managers and customer to validate their requirements. Moreover, the requirements cannot be elicited entirely at one time. After validating the requirements, the customers and product managers would fix the faults of the requirements and find new interesting functional requirements of the target system. This fixing and eliciting requirements process will make the previous requirements version from $R_{n-1}$ to next version $R_n$ in Figure \ref{refinement}. Then the evolved requirement $R_n$ will be used to generate a new prototype $P_n$. Compared with the methods to manually implementing the prototype, customers and project managers can use RM2PT to generate new prototypes and validate the requirements without waiting for the software engineers to implement the prototype. Therefore, RM2PT shortens the timeline and boosts the software development process. Moreover, RM2PT reduces the inconsistency between the requirements and the prototype by automatically transformation rules; otherwise, errors may be involved by the green hands or careless of the human nature. In the next section, we will demonstrate the validity of the RM2PT through four classical case studies using in our daily life.

\section{Case Studies}
In this section, we use four case studies to demonstrate the validity and capacity of the proposed approach for requirements validation and evolution. Those case studies are wildly used systems in our daily life: supermarket system (CoCoME), Library Management System (LibMS), Automated Teller Machine (ATM), and Loan Processing System (LoanPS). CoCoME mainly contains the scenarios of the cashier to process sales in supermarkets, and supermarket managers to manage the storage of products. LibMS primarily involves the student to borrow and return books. ATM concerns the customer to withdraw and credit money. LoanPS touches the use cases of submitting and evaluating loan requests, booking a new loan and making the payment. Those four case studies demonstrate the different aspects of requirements modeling and prototyping of RM2PT. ATM provides a quick-start and simple demonstration of RM2PT about requirements modeling and prototyping. CoCoME demonstrates the capacity of generation and validation complex use case such as \textit{processSale}. LibMS demonstrates the capacity of generation and validation of the complex system operations, that makes the case study of LibMS contains the highest number of primitive operations. The last one LoanPS demonstrates the ability to invoking third-party system services by including the remote operations calling from the account management system of banks and the credit management system of the governments. We will show the requirements modeling and validation results in the following parts. More details of the requirements models can be found at GitHub\footnote{\url{https://github.com/RM2PT/CaseStudies}}.

\subsection{Evaluation Complexity of Requirements Model}
Requirements complexity is measured by the dimensions of the number of the actors, use cases, system operations, entity classes, and associations of the entity classes in the requirements model. We present the requirements complexity of four case studies in Table \ref{complexity}. 
\begin{table}[!htb]
\ra{1.2}
  \caption{The Complexity of Requirements Models}
  \label{complexity}
  \centering
  \begin{adjustbox}{max width=\columnwidth}
  \begin{threeparttable}
  \begin{tabular}{lccccccc}
    \toprule
    Case Study & Actor & Use Case & SO & AO & Entity Class & Association & INV \\
    \midrule
    ATM & 2 & 6 & 15 & 103 & 3 & 4 & 5 \\
    CoCoME & 3 & 16 & 43 & 273 & 13 & 20 & 10 \\
    LibMS & 7 & 19 & 45 & 433 & 11 & 17 & 25 \\
    LoanPS & 5 & 10 & 34 & 171 & 12 & 8 & 12 \\
    \midrule
    Sum & 17 & 51 & 137 & 980  & 39 & 49 & 52 \\
   \bottomrule
  \end{tabular}
  
  \begin{tablenotes}[para]
   \footnotesize
         \item[*] Above table shows the number of elements in the requirements model. SO and AO are the abbreviations of system and primitive operations responsibility respectively. INV is the abbreviation of invariant.\\
   \end{tablenotes}
  
  \end{threeparttable}
  \end{adjustbox}
\end{table}
The requirements model of CoCoME contains three actors, sixteen use cases, forty-three system operations, two hundred-seven-three primitive operations, thirteen entity classes, twenty associations between those objects, and ten invariants. The requirements model of LibMS includes seven actors, nineteen use cases, forty-five system operations, four hundred-thirty-three primitive operations, eleven entity classes, seventeen associations, and twenty-five invariants. The requirements model of ATM includes two actors, six use cases, fifteen system operations, one hundred and three primitive operations, three entity classes, four associations, and five invariants. The requirements model of LoanPS includes five actors, ten use cases, thirty-four system operations, one hundred and seventy-one three primitive operations, twelve entity classes, eight associations, and twelve invariants. In short, we provides \textbf{17} actors, \textbf{51} use cases, \textbf{137} system operations, \textbf{980} primitive operation, \textbf{39} entity classes, \textbf{49} associations of entity classes, and \textbf{52} invariants. 


\subsection{Results of Requirements Modeling and Generation}
The generation results of four case studies are shown in this subsection. The failures of requirements modeling and generation can be divided into two situations: 1) the contracts of system operations cannot be correctly specified in OCL expression without invoking third-party APIs such as sorting algorithm, sending email, and etc. 2) the contracts of system operations can be correctly specified in OCL expression, but no transformation rule is matched to generate the implementation for the system operations correctly. We count both failures of those case studies by the measurements \textit{MSuccess} and \textit{GenSuceess} in Table \ref{GenereationAndValidation}. 

\begin{table}[!htb]
\ra{1.2}
  \caption{The Generation Result of System Operations}
  \label{GenereationAndValidation}
  \centering
  \begin{adjustbox}{max width=1\columnwidth}
  \begin{threeparttable}
  \begin{tabular}{lcccc}
    \toprule
    Case Study & NumSO & MSuccess & GenSuccess & SuccessRate (\%) \\
    \midrule
   ATM & 15 & 15 & 15 & 100 \\
   CoCoME & 43 & 41 & 40 & 93.02  \\
   LibMS & 45 & 43 & 42 & 93.33\\
   LoanPS & 34 & 30 & 30 & 88.23 \\
   \midrule
   Average & 34.25 & 32.25 & 31.75 & 93.65 \\
   \bottomrule
  \end{tabular}
  
  \begin{tablenotes}[para]
   \footnotesize
         \item[*] MSuccess is the number of SO which is modeled correctly without external event-call, GenSuccess is the number of SO which is successfully generated, SuccessRate = GenSuccess / NumSO.
   \end{tablenotes}
  
  \end{threeparttable}
  \end{adjustbox}
  
\end{table}

\noindent \textbf{ATM} is the most straightforward case study of four. It does not contain a complex workflow. All of the fifteen system operations can be successfully modeled and generated. The successful generation rate reached 100\%. 

\vspace{.1cm}
\noindent \textbf{CoCoME} contains forty-three system operations. Forty-one can be successfully generated, in which two system operations cannot be specified correctly. The two failures of requirements modeling are 1) listing top 10 out of stock products, 2) send emailing notification to the student, who holds the book copy will be due in the next two days. Preconditions of those two system operations can be specified correctly, but the sub-expressions of post-condition, \textit{listing top 10} and \textit{sending email}, can not be specified correctly in OCL expression. Therefore, the corresponding transformation and implementation cannot be generated correctly as well. 

Another case is that listing almost out of stock the products with storage less than 10 in ascending order. That post-condition can be specified correctly, but the sub-expression of \textit{ascending order} cannot be generated because no transformation rule can be used to map that sub-expression to the primitive operations. Counting the failures of system operations modeling and generation, the successful rate of prototype \textit{CoCoME} is 93.02\%.

\vspace{.2cm}
\noindent \textbf{LibMS} contains forty-five system operations, forty-three can be correctly modeled, and forty can be correctly generated. The two modeling failures are 1) listing the top 10 the student holding the overdue book over a week. 2) sending an email notification to the student, who holds the book copy will be due in the next two days.  Like CoCoMe, the failures of modeling are also due to the expression \textit{top 10} and \textit{sending email}. The failure of generation is listing the holding book records by the due day in \textit{ascending order}. Counting in those failures, the successful generation rate is 91.11\%.

\vspace{.2cm}
\noindent \textbf{LoanPS} contains thirty-four system operations, and thirty system operations can be correctly specified and generated. Four failures of modeling include 1) listing the top 10 loan requests with loan assistant, 2) sending a notification to the applicant when her loan request is approved, 3) printing the loan agreement, 4) sending the notification to the customer when their loan is due soon. The successful generation rate of \textit{LoanPS} is 88.23\%.

In short, RM2PT can successfully generate average \textbf{93.65\%} system operations of those four case studies without any extension. There still has \textbf{6.35\% errors} of requirements modeling and prototype generation. As we mentioned, the errors mainly caused by the failures of specifying the post-conditions of the contract such as sending email, printing document, sorting and retrieving top N elements. 


To help users easier finding the failures of modeling and prototype generation, 1) our case tool RM2PT can automatically indicate the sub-expression error with red underlines in the post-conditions when generating the prototype from the requirement model and the code errors in the generated prototype. 2) RM2PT provides an extension mechanism to allow specifying the third-party APIs were invoking in OCL expression. Once errors were identified by RM2PT, we can replace the error expression in post-condition with the calling expression of the third-party service. This calling expression will be a transformation to the operation calling in the prototype. The developer can manually implement the operation, or invoking the operation system APIs (for sending email and printing document) and reusing the collection algorithm library (for sorting and top N elements retrieving). After specified the calling expression of third-services into the requirements model, all the system operations of four case studies can be correctly modeled and generated. 

This subsection exposes the results of requirements modeling, prototype generation, and how to fix the failures of modeling and generation. Based on the techniques of requirements validation and evolution in the section \ref{sec:requirementvalidation}, we will show the result of validation and evolution in next subsection.

\subsection{Results of Requirements Validation and Evolution}
The result of requirements validation and evolution of four case studies are shown in Table \ref{RequirementsErrors} and \ref{RequirementsMissing}, which contain the statistic of requirements errors and missing.
\begin{table}[!htb]
\ra{1.2}
  \caption{Requirements Errors}
  \label{RequirementsErrors}
  \centering
  \begin{adjustbox}{max width=1\columnwidth}
  \begin{threeparttable}
  \begin{tabular}{lccc}
    \toprule
    & \multicolumn{2}{c}{Requirements Errors}
    \\ \cmidrule(lr){2-3} 
    Name & Pre-condition & Post-condition \\
    \midrule
   ATM & 5 & 12  \\
   CoCoME & 8 & 23   \\
   LibMS & 12 & 26  \\
   LoanPS & 6 & 21 \\
   \midrule
   Total & 31 & 68  \\
   \bottomrule
  \end{tabular}
  
  \begin{tablenotes}[para]
   \footnotesize
   \end{tablenotes}
  
  \end{threeparttable}
  \end{adjustbox}
  
\end{table}
During the three-round modeling, prototype generation, requirements validation, we found \textbf{99} requirements errors, which includes 31 errors in the pre-condition and 68 errors in the post-condition. In details, 5 pre-condition and 12 post-condition errors are founded in ATM. 8 pre-condition and 23 post-condition are founded in CoCoME case study. 12 pre-condition and 26 post-condition errors are founded in LibMS case study. 6 pre-condition and 21 post-condition are founded in LoanPS case study.

\begin{table}[!htb]
\ra{1.2}
  \caption{Requirements Missing}
  \label{RequirementsMissing}
  \centering
  \begin{adjustbox}{max width=1\columnwidth}
  \begin{threeparttable}
  \begin{tabular}{lcccccc}
    \toprule
    & \multicolumn{5}{c}{Requirements Missing} \\ 
    \cmidrule(lr){2-7}  
    Name & Actor & UseCase & SO & Entity Class & Association & INV\\
    \midrule
   ATM & 1 & 3 & 9 & 1 & 2 & 3\\
   CoCoME & 1 & 11 & 22 & 5 & 10 & 5 \\
   LibMS & 4 & 12 & 14 & 11 & 15 & 12\\
   LoanPS & 2 & 3 & 15 & 4 & 2 & 8\\
   \midrule
   Total & 8 & 29 & 60 & 21 & 29 & 28\\
   \bottomrule
  \end{tabular}
  
  \begin{tablenotes}[para]
   \footnotesize
   \end{tablenotes}
  
  \end{threeparttable}
  \end{adjustbox}
  
\end{table}
Requirement validation can help to find errors in requirements, but also can help to find missing requirements. During the requirement validations and refinements from version one the version three, \textit{ATM} add one actor, three use cases, nine system operations, one entity class, two associations of entity classes, and two invariants. CoCoME adds one actor, eleven use cases, twenty-two system operations, five entity classes, ten associations of entity classes, and five invariants. LibMS adds four actors, twelve use cases, fourteen system operations, eleven entity classes, ten associations of entity classes, and twelve invariants. LoanPS adds two actors, three use cases, fifteen system operations, four entity classes, two associations of entity classes, and eight invariants. Totally, we find \textbf{173} missing requirements during the requirements validations and refinements about \textbf{6} actors, \textbf{29} use cases, \textbf{60} system operations, \textbf{21} entity classes, \textbf{29} the associations of entity classes, and \textbf{28} invariants. 


\subsection{Performance of Prototype Generation}
The advantages of the proposed approach to automated generate prototype are more efficient than the manual prototype implementation by the developers. We evaluate the RM2PT prototyping performance by comprising prototype time among RM2PT and the developers. The performance is measured on the dimensions: a) line of code, which is calculus by the tool \textit{cloc}\footnote{\url{https://github.com/AlDanial/cloc}}, b) prototype time of RM2PT, c) generation time for the system operations, d) average coding time of ten master students from computer science, e) average coding time of ten developers from the industry with at least three year coding experiments. All the measurement are computed on the PC with 3.5 GHz Intel Core i5, 16 GB 1600 MHz DDR3, and 500 GB Flash Storage.

\begin{table}[!htb]
\ra{1.2}
  \caption{Prototyping Performance}
  \label{prototypingperformance}
  \centering
  \begin{adjustbox}{max width=1\columnwidth}
  \begin{threeparttable}
  \begin{tabular}{lccccc}
    \toprule
   Name & Line of Code & Prototype Time (ms) & SO Time(ms) & Student (hour) & Developer (hour) \\
    \midrule
   ATM & 3897 & 309.74 & 2.26 & 8.3 & 3.2\\
   CoCoME & 9572 & 788.99 & 9.78 & 19.2 & 10.6\\
   LibMS & 12017 & 1443.39 & 18.22 & 25.1 & 14.3\\
   LoanPS & 7814 & 832.78 & 5.52 & 16.2 & 8.4\\
    \midrule
   Average & 8325 & 843.73 & 8.95 & 17.20 & 9.14 \\
   \bottomrule
  \end{tabular}
  
  \begin{tablenotes}[para]
   \footnotesize
   \end{tablenotes}
  
  \end{threeparttable}
  \end{adjustbox}
  
\end{table}

The comparison result is presented in Table \ref{prototypingperformance}. ATM contains 3897 lines of code, the generation of prototype spends 309.74 ms, in which system operations only spend 2.26 ms, the average coding time of students is 8.3 hours, but developers require 3.2 hours. CoCoME contains 9572 lines of code, generation prototype spend 788.99 ms, the average coding time of students is 19.2 hours, developers need 10.6 hours. LibMS includes 12017 lines of code, generation times of prototype is 1443.39 ms, the average coding time of students is 25.1 hours, and developers only spend 14.3 hours. LoanPS contains 7814 lines of code, the generation time of prototype is 832.78 ms, the average coding time of student is 16.2 hours, and the developers require only 8.4 hours. In average, the case studies contains 8325 lines of code, generation prototype spend \textbf{843.73 ms} less than 1 second, the system operations generations only require \textbf{8.95 ms}, students need \textbf{17.20 hours} and experienced developers require \textbf{9.14 hours}.

RM2PT auto-prototyping is much more efficient than manual prototyping (\textbf{\textasciitilde 1 second vs \textasciitilde 9 hours}). Although the experienced developers take less prototyping time than the students, that is still much in-efficient than auto-prototyping approach. Moreover, the manual prototyping usually introduces inconsistency between prototype and requirements because of careless and not understanding of requirement model. That makes more than half of the whole prototyping time to debugging. Furthermore, if we only count the time for fulfillment of system operations, the spending time \textbf{\textasciitilde 9 ms)} is much less than the prototyping (\textbf{\textasciitilde 850 ms}). In short, RM2PT is an efficient and effective approach to generate prototype without introducing inconsistency between the requirements model and prototype.

\subsection{Limitation}
The case studies demonstrate the effective and efficient of RM2PT for requirements modeling, prototype generation, requirements validation, and evolution. However, it has some limitations. 1) The first one is that 6.91\% system operations cannot be correctly specified or successfully generated without introducing third-party services, but this limitation has been solved by specified the thrid-part service is invoking in OCL expression. That is a significant extension of our previous work. 2) The second limitation that is although the formal specification OCL has short leaning cure than other formal specification, it still needs time for learning to specify the correct contract. We will try to find the better solution to alleviate this problem. 3) The third limitation is the generation performance, and it can be further optimized in the future.

\section{Related Work}


The related work contains three parts: 1) automated prototype approach, 2) requirements modelers, and 3) the comparison with our previous work.


%
%
%
\subsection{Automated Prototype Generation}
The following approaches are the most closely related to our work. Umple \cite{Forward2012} can generate a prototype from a class model (conceptual class diagram) and state machine models. However, the state machine only contains abstract states and descriptions of actions (system operations). To generate fully functional prototypes, the actions must be implemented by programming languages. Moreover, invariant checking and requirements validation are not touched. 

ActionGUI \cite{ActionGUI2014} can generate a multi-tier application from a design model, which includes a data model (specified by ComponentUML \cite{Basin:2006:MDS:1125808.1125810}), a security model (specified by SecurityUML \cite{Basin:2006:MDS:1125808.1125810}) and a GUI model (specified by GUI Modelling Language). Compared with our approach, there are three main differences. ActionGUI a) requires to provide system operation design by using data actions (primitive operations) to specify statements for events (system operations) in GUI model. A statement (the workflow of primitive operations) is either an action, an iteration or a sequence of statements. Our approach only requires providing contracts (pre and post conditions in OCL) to system operations, then the implementation of system operation can be automated generated. b) ActionGUI requires GUI engineer to construct a GUI model through their specific tool manually. Our approach can be automated generate prototype from requirements model without providing any GUI design. c) the generated GUI of our approach is implemented by state-of-arts GUI architecture design, i.e., separated concerns into GUI content (XML) and GUI style (CSS). That makes prototype high reusable by only applying user-friendly styles to the same content.

The paper \cite{Elkoutbi2006} proposed an intermediate approach to generate UI prototypes from UML models. It generates state chart diagrams from a design model specified by a class diagram and collaborations diagrams for each use case, and then generate the UI prototype from a use case diagram and the intermediate state chart diagrams. Comparison with our work, we only require a requirements model, which includes a use case diagram, conceptual class diagram (without system operations), system sequence diagrams and the contracts of system operations (only including the interaction between actors and system interface without internal interactions among objects, that required in collaborations diagrams). The generated prototype contains mechanisms for pre-conditions and invariants checking.

JEM \cite{PASTOR2001507} can generate an n-tiered prototype from a conceptual model and an execution model. The conceptual model contains a) an object model specified in a class diagram, b) functional models for the attributes of the class and c) dynamic model for each class defined in state chart diagram. Then the prototype can be generated by the derived formal specification OASIS from the conceptual model and the implementation related execution model, which includes the generation templates and details mapping about OASIS to the implementation. Compared with our approach, pre- and post-conditions only contain simple attribute checking and updating (only one attribute involved) that is hard for working on the practical requirements model and prototype generation. 

SCORES \cite{Homrighausen2002} proposed a semi-automatically approach to generating prototypes from an enhancement of the requirements specification with user interface model in FLUID \cite{Kosters1996}. Requirements specification contains a use case diagram, activities diagrams to each use case, and a class diagram (including operations). User interface model includes a specification for view widgets, their navigation, and selection or manipulations (primitive operations) of the domain objects. It does not include the specification or contract for system operations other than simple manipulations of domain objects. Therefore, the sophisticated system operations cannot be generated such as \textit{borrowBook}, which includes collaborations of primitive operations such as find an object, form a link, update the attribute. Moreover, the class diagram in SCORES already contains the system operations in the activities diagram, stickily speaking, that is a design model.

In short, the related works 1) require providing a design model, which contains a class diagram encapsulating system operations, the design or implementation of system operations specified in collaboration diagrams or a programming language. Moreover, SCORES, ActionGUI and JEM require a GUI design for generating prototype user interface. 2) They lack evaluations about the prototype generation, the mechanism to deal with the non-executable elements in the requirements model. 3) Requirements validation and evolution are not touched. The generated prototype does not provide mechanisms to invariants checking and object status observations in run-time for requirements validation and evolution.

\subsection{Requirements Modelers}
%
%
%
Most UML modeling tools support OCL-based contract and can generate skeleton code for entity classes in the conceptual class model. Moreover, Visual Paradigm (VP) \cite{Paradigm2013} not only supports entity class generation but also supports generating code for primitive operations of entity class such as creating, deleting, modifying, finding the entity object. And the generated entity classes can be automated mapped into tables of relational database thought ORM and the corresponding RESTFul-based web service wrappers could be automatically generated. Eclipse Foundation provides many open source CASE tools and frameworks. For example, Papyrus UML\cite{Lanusse2009} is well developed and widely used open source tool. EMF Forms\footnote{\url{http://www.eclipse.org/ecp/emfforms}} cannot only generate primitive operations (like in VP) but also can generate GUI for validating those operations. CDO\footnote{\url{http://www.eclipse.org/cdo/}} provides open source solutions to support ORM for EMF\footnote{\url{https://eclipse.org/modeling/emf/}} model. Commercial CASE tool Enterprise Architect\footnote{\url{http://www.sparxsystems.com}} supports generate system operations of objects from the presented design model. The study \cite{Kundu2013} proposed sequence integration graph (SIG), which acts as an intermediate to help automatically fulfill system operations from sequence diagram. Morever, if providing design model, MasterCraft \cite{kulkarni2002generating} could generate information system prototype. AndroMDA\footnote{\url{http://andromda.sourceforge.net}} could generate Java EE and .NET system prototype. However, all the current CASE tools can not generate a prototype without providing an explicit design model. USE (UML-based Specification Environment) \cite{gogolla2007use} supports analysts, designers, and developers in executing UML models and checking OCL constraints and thus enables them to employ model-driven techniques for software production. The project of Eclipse OCL\footnote{\url{https://projects.eclipse.org/projects/modeling.mdt.ocl}} provides an OCL parser and evaluator on UML models. Although USE and Eclipse OCL can evaluate the invariants and precondition of operation contracts, they can not generate the implementation of system operation to conform the post-condition.

\subsection{Compared with Our Previous Work}
%
%
%
Our previous works about Refinement of Component and Object Systems (rCOS) modeler \cite{Liu2005}\cite{Jifeng2006}\cite{Ke2012} and Automated Prototype Generation and Analysis (AutoPA) \cite{Li2008}\cite{Li2010} touch object-oriented modeling and prototype generation. The rCOS supports refinement calculus for both component-based and object-oriented models. It uses first-order logic to specify the contracts of system operations. It focuses on modeling, but can not generate the prototype. AutoPA can generate system operations from the requirements model with OCL contracts to demonstrate the feasibility of our approach. Compared with our previous works \cite{Li2008}\cite{Li2010}, the proposed approach and implemented RM2PT provides:

\vspace{.1cm}
\noindent 1) \textit{Extensions of transformation rules.} Our previous work only cover a small subset of OCL expression that describes creating an object, adding a link, getting the value of attribute, setting an attribute, removing a link, and deleting an object, but not includes: a) finding an object with a condition such as \textit{any()} and \textit{select()}, b) OCL standard operations such as \textit{size()} and \textit{isEmpty()}, c) the iteration expression \textit{forAll()} for iterating object from a list, d) invoking third-party operations, and e) the \textit{return} expression.

In details, we introduce the rules from $R_{1}$ to $R_{3}$ to find the object with a condition through \textit{EntityManager}, which stores all the references of objects. Moreover, the rules from $R_4$ to $R_7$ support finding the objects through the links. The rules from $R_{8}$ to $R_{11}$ support OCL standard operations. The checking of an object exists (or not exists) can be fulfilled by the combination of the rules $R_{8}$ and $R_{13}$ (or $R_{14}$), and checking a link exists that is supported by the composition of the rules $R_4$, $R_5$, and $R_8$. Besides, the uniqueness of an object can be checked through the rule $R_{15}$. The getting and setting an attribute of object is the same as the rules $R_{12}$ and $R_{23}$. The rule of creating an object is refined to two rules, the rule $R_{16}$ is to create an object, and the rule $R_{17}$ is to add the created objects into the system (add the reference to the created object into \textit{EntityManager}). The rule of releasing an object is the same as the rules $R_{18}$. We refine the rules of adding a link to two rules: the first rule $R_{19}$ is adding a link of the one-to-one association, the second rule $R_{21}$ is for the one-to-many association. Likewise, the rules of removing a link is refined to the rules $R_{20}$ and $R_{22}$. More general system operation such as all the objects of a class satisfy a particular constraint are common requirements in a contract (E.g., when time passed 00:00 AM, the attributes of remain days of all the students were equaled to the previous value plus one.). We add the rule $R_{24}$ for iteration expression, the rule $R_{25}$ for the \textit{return} expression, and the rule $R_{26}$ for invoking third-party APIs library.

\vspace{.1cm}
\noindent 2) \textit{Supporting requirements executable analysis and wrapping the non-executable specification into an interface}. Our previous works face the problem of un-executable elements of the contract of system operation, which cannot be fulfilled by primitive operations, like sorting and filtering. RM2PT provides a mechanism to identify and un-executable parts of contract and wrapper them into an interface, which can be fulfilled by third-party APIs.

\vspace{.1cm}
\noindent 3) \textit{Requirements validation and evolution through the generated prototypes.} Our previous work does not touch how to use the generated prototypes to requirements validation and evolution. In this paper, we provide mechanisms to observe the state of objects, and pre-condition and invariants checking in the prototype and show how to validate and evolve requirements through the generated prototype from RM2PT. Four case studies from different aspects demonstrate effective and efficient of our proposed approach. 

\vspace{.1cm}
\noindent 4) \textit{Self-contained requirement modeler and OCL parser}. RM2PT does not rely on third-party UML tools and OCL parser, RM2PT self-contains a requirement modeler and an OCL parser with bi-directionally synchronization between graphics and textual requirements models. The graphics model provides the easer-understood visualized notations for communication with customers. Textual requirement model makes developers easily define the contracts of responsibility. Furthermore, by integrating with version control tools such as SVN and Git, RM2PT can support the trackable iteration for evolutionary requirements elicitation. 


\vspace{.1cm}
\noindent 5) \textit{Generated prototypes based on architecture and design patterns under GRASP guidelines and design patterns.} Unlike AutoPA encapsulates all the system operations into conceptual classes, RM2PT separates concerns into a different abstract level. After generating supporting classes related to architecture, RM2PT automatically encapsulate primitive operations into entity classes and system operations into fabricated classes for separating concerns. Furthermore, different responsibilities of holding content and representation are separated through the latest Java GUI framework JavaFX. The generated prototype takes the architecture and design patterns similar to Java EE and .NET enterprise system, which is capable of extending for a practical scenario. 


\section{Conclusion and Future Work}

This paper presents an approach to automated prototype generation from a formal requirements model, and the requirements model can be validated and evolved by the generated prototype. It includes executable analysis of formal specification and designs a set of transformation rules for translating the executable parts of the contract into Java source code. The non-executable parts of contract can be identified and wrapped by an interface, which can be fulfilled by third-party APIs. Four cases studies, which are library management system, ATM, CoCoME and loan processing system, have been investigated, and the experiment result is satisfactory that the \textbf{93\%} of use cases can be generated successfully. The CASE tool: RM2PT and its tutorials are available for the public at GitHub\footnote{\url{http://rm2pt.mydreamy.net}}. 

In the future, we will improve the current transformation algorithm to cover the more substantial subset of the executable specification. Meanwhile, we will integrate current prototyping tool with our another work on automated translating use case definitions in natural language into their corresponding formal contract in OCL. That will make this work more applicable to software industrial developers. Generally, they can read formal specification, but they have difficulties in writing a formal specification. With the tool support, their task is to confirm whether the translated formal specification is conformance with the natural language requirement description. Furthermore, after a system requirements model is validated by prototyping, we can generate the prototype into its corresponding real system with another developed transformation software tool. Besides, we will investigate how to generate test cases from the OCL specification, so that we can enhance our tool for automated prototyping and testing. Finally, the tool can be used and checked with more case studies, and hopefully, it can benefit the software industry during requirements engineering.

\ifCLASSOPTIONcompsoc
  \section*{Acknowledgments}
\else
  \section*{Acknowledgment}
\fi

This work was supported by Macau Science and Technology Development Fund (FDCT) (No. 103/2015/A3) and University of Macau (No. MYRG 2017-00141-FST), 1000-Expert Program Grant (No. SWU116007), and National Natural Science Foundation of China (NSFC) (No. 61472779, 61562011, 61672435 and 61732019)

\ifCLASSOPTIONcaptionsoff
  \newpage
\fi



\bibliographystyle{IEEEtran}
\bibliography{IEEEabrv}

\begin{thebibliography}{10}
\providecommand{\url}[1]{#1}
\csname url@samestyle\endcsname
\providecommand{\newblock}{\relax}
\providecommand{\bibinfo}[2]{#2}
\providecommand{\BIBentrySTDinterwordspacing}{\spaceskip=0pt\relax}
\providecommand{\BIBentryALTinterwordstretchfactor}{4}
\providecommand{\BIBentryALTinterwordspacing}{\spaceskip=\fontdimen2\font plus
\BIBentryALTinterwordstretchfactor\fontdimen3\font minus
  \fontdimen4\font\relax}
\providecommand{\BIBforeignlanguage}[2]{{%
\expandafter\ifx\csname l@#1\endcsname\relax
\typeout{** WARNING: IEEEtran.bst: No hyphenation pattern has been}%
\typeout{** loaded for the language `#1'. Using the pattern for}%
\typeout{** the default language instead.}%
\else
\language=\csname l@#1\endcsname
\fi
#2}}
\providecommand{\BIBdecl}{\relax}
\BIBdecl

\bibitem{DBLP:journals/re/SutcliffeEM99}
A.~G. Sutcliffe, A.~Economou, and P.~Markis, ``Tracing requirements errors to
  problems in the requirements engineering process,'' \emph{Requirements
  Engineering}, vol.~4, no.~3, pp. 134--151, 1999.

\bibitem{Hofmann2001}
H.~F. Hofmann and F.~Lehner, ``Requirements engineering as a success factor in
  software projects,'' \emph{IEEE Software}, vol.~18, no.~4, pp. 58--66, Jul.
  2001.

\bibitem{sommerville2015software}
I.~Sommerville, \emph{Software Engineering}.\hskip 1em plus 0.5em minus
  0.4em\relax Addison Wesley, 2015.

\bibitem{Atladottir2012}
G.~Atladottir, E.~T. Hvannberg, and S.~Gunnarsdottir, ``Comparing task
  practicing and prototype fidelities when applying scenario acting to elicit
  requirements,'' \emph{Requirements Engineering}, vol.~17, no.~3, pp.
  157--170, Sep. 2012.

\bibitem{Lichter1994}
H.~Lichter, M.~Schneider-Hufschmidt, and H.~Zullighoven, ``Prototyping in
  industrial software projects-bridging the gap between theory and practice,''
  \emph{IEEE Transactions on Software Engineering}, vol.~20, no.~11, pp.
  825--832, Nov. 1994.

\bibitem{Kordon2002}
F.~Kordon and Luqi, ``An introduction to rapid system prototyping,'' \emph{IEEE
  Transactions on Software Engineering}, vol.~28, no.~9, pp. 817--821, Sep.
  2002.

\bibitem{userinterface}
D.~Baumer, W.~Bischofberger, H.~Lichter, and H.~Zullighoven, ``User interface
  prototyping-concepts, tools, and experience,'' in \emph{Proceedings of IEEE
  18th International Conference on Software Engineering (ICSE'96)}, Mar. 1996,
  pp. 532--541.

\bibitem{Kamalrudin2011}
M.~Kamalrudin and J.~Grundy, ``Generating essential user interface prototypes
  to validate requirements,'' in \emph{Proceedings of the 26th IEEE/ACM
  International Conference on Automated Software Engineering (ASE'11)}, Nov.
  2011, pp. 564--567.

\bibitem{Ciccozzi2018}
F.~Ciccozzi, I.~Malavolta, and B.~Selic, ``Execution of {UML} models: a
  systematic review of research and practice,'' \emph{Software and Systems
  Modeling}, Apr. 2018.

\bibitem{Regep2000}
D.~Regep and F.~Kordon, ``Using metascribe to prototype a {UML} to {C++/Ada95}
  code generator,'' in \emph{Proceedings of the 11th International Workshop on
  Rapid System Prototyping (RSP'00)}, 2000, pp. 128--133.

\bibitem{Kundu2013}
D.~Kundu, D.~Samanta, and R.~Mall, ``Automatic code generation from unified
  modelling language sequence diagrams,'' \emph{IET Software}, vol.~7, no.~1,
  pp. 12--28, 2013.

\bibitem{kulkarni2002generating}
V.~Kulkarni, R.~Venkatesh, and S.~Reddy, ``Generating enterprise applications
  from models,'' in \emph{Proceedings of the 8th International Conference on
  Object-Oriented Information Systems (OOIS'02)}.\hskip 1em plus 0.5em minus
  0.4em\relax Springer, 2002, pp. 270--279.

\bibitem{DBLP:conf/compsac/LiLH01}
X.~Li, Z.~Liu, and J.~He, ``Formal and use-case driven requirement analysis in
  {UML},'' in \emph{Proceedings of the 25th International Computer Software and
  Applications Conference (COMPSAC'01)}, Chicago IL, USA, 2001, pp. 215--224.

\bibitem{zhiming2003}
Z.~Liu, H.~Jifeng, X.~Li, and Y.~Chen, ``A relational model for formal
  object-oriented requirement analysis in {UML},'' in \emph{Proceedings of the
  5th International Conference on Formal Engineering Methods (ICFEM'2003)},
  Berlin, Heidelberg, 2003, pp. 641--664.

\bibitem{DBLP:journals/scp/ChenLRSZ09}
Z.~Chen, Z.~Liu, A.~P. Ravn, V.~Stolz, and N.~Zhan, ``Refinement and
  verification in component-based model-driven design,'' \emph{Science of
  Computer Programming}, vol.~74, no.~4, pp. 168--196, 2009.

\bibitem{liu2002object}
Z.~Liu, ``Object-oriented software development in {UML},'' UNU/IIST, Macau,
  Tech. Rep., 2002.

\bibitem{meyer2002design}
B.~Meyer, \emph{Design by contract}.\hskip 1em plus 0.5em minus 0.4em\relax
  Prentice Hall, 2002.

\bibitem{Li2008}
D.~Li, X.~Li, J.~Liu, and Z.~Liu, ``Validation of requirement models by
  automatic prototyping,'' \emph{Innovations in systems and software
  engineering}, vol.~4, no.~3, pp. 241--248, 2008.

\bibitem{Li2010}
X.~Li, Z.~Liu, M.~Schaf, and L.~Yin, ``{AutoPA}: automatic prototyping from
  requirements,'' in \emph{Leveraging Applications of Formal Methods,
  Verification, and Validation}.\hskip 1em plus 0.5em minus 0.4em\relax
  Springer, 2010, pp. 609--624.

\bibitem{alexander1977pattern}
C.~Alexander, S.~Ishikawa, M.~Silverstein, J.~R. i~Rami{\'o}, M.~Jacobson, and
  I.~Fiksdahl-King, \emph{A pattern language}.\hskip 1em plus 0.5em minus
  0.4em\relax Gustavo Gili, 1977.

\bibitem{vlissides1995design}
J.~Vlissides, R.~Helm, R.~Johnson, and E.~Gamma, \emph{Design patterns:
  Elements of reusable object-oriented software}.\hskip 1em plus 0.5em minus
  0.4em\relax Addison-Wesley, 1995.

\bibitem{richards2015software}
M.~Richards, \emph{Software Architecture Patterns}.\hskip 1em plus 0.5em minus
  0.4em\relax O’Reilly, Sebastopol, 2015.

\bibitem{Fowler2002}
M.~Fowler, \emph{Patterns of enterprise application architecture}.\hskip 1em
  plus 0.5em minus 0.4em\relax Addison-Wesley, 2002.

\bibitem{Krasner:1988:CUM:50757.50759}
G.~E. Krasner and S.~T. Pope, ``A cookbook for using the model-view controller
  user interface paradigm in {Smalltalk-80},'' \emph{Journal of Object-Oriented
  Programming}, vol.~1, no.~3, pp. 26--49, Aug. 1988.

\bibitem{bettini2016implementing}
L.~Bettini, \emph{Implementing domain-specific languages with Xtext and
  Xtend}.\hskip 1em plus 0.5em minus 0.4em\relax Packt Publishing Ltd, 2016.

\bibitem{Forward2012}
A.~Forward, O.~Badreddin, T.~C. Lethbridge, and J.~Solano, ``Model-driven rapid
  prototyping with umple,'' \emph{Software: Practice and Experience}, vol.~42,
  no.~7, pp. 781--797, Jul. 2012.

\bibitem{ActionGUI2014}
D.~Basin, M.~Clavel, M.~Egea, M.~A.~G. de~Dios, and C.~Dania, ``A model-driven
  methodology for developing secure data-management applications,'' \emph{IEEE
  Transactions on Software Engineering}, vol.~40, no.~4, pp. 324--337, Apr.
  2014.

\bibitem{Basin:2006:MDS:1125808.1125810}
D.~Basin, J.~Doser, and T.~Lodderstedt, ``Model driven security: From {UML}
  models to access control infrastructures,'' \emph{ACM Transactions on
  Software Engineering and Methodology}, vol.~15, no.~1, pp. 39--91, Jan. 2006.

\bibitem{Elkoutbi2006}
M.~Elkoutbi, I.~Khriss, and R.~K. Keller, ``Automated prototyping of user
  interfaces based on {UML} scenarios,'' \emph{Automated Software Engineering},
  vol.~13, no.~1, pp. 5--40, Jan. 2006.

\bibitem{PASTOR2001507}
O.~Pastor, J.~Gómez, E.~Insfrán, and V.~Pelechano, ``The {OO-method} approach
  for information systems modeling: from object-oriented conceptual modeling to
  automated programming,'' \emph{Information Systems}, vol.~26, no.~7, pp. 507
  -- 534, 2001.

\bibitem{Homrighausen2002}
A.~Homrighausen, H.-W. Six, and M.~Winter, ``Round-trip prototyping based on
  integrated functional and user interface requirements specifications,''
  \emph{Requirements Engineering}, vol.~7, no.~1, pp. 34--45, Apr. 2002.

\bibitem{Kosters1996}
G.~Kosters, H.~W. Six, and J.~Voss, ``Combined analysis of user interface and
  domain requirements,'' in \emph{Proceedings of the 2th International
  Conference on Requirements Engineering (RE'96)}, Apr. 1996, pp. 199--207.

\bibitem{Paradigm2013}
V.~Paradigm, ``Visual paradigm for {UML},'' \emph{Visual Paradigm for UML - UML
  tool for software application development}, 2013.

\bibitem{Lanusse2009}
A.~Lanusse, Y.~Tanguy, H.~Espinoza, C.~Mraidha, S.~Gerard, P.~Tessier,
  R.~Schnekenburger, H.~Dubois, and F.~Terrier, ``Papyrus {UML}: an open source
  toolset for mda,'' in \emph{Proceedings of the 5th European Conference on
  Model-Driven Architecture Foundations and Applications (ECMDA-FA'09)}.\hskip
  1em plus 0.5em minus 0.4em\relax Citeseer, 2009, pp. 1--4.

\bibitem{gogolla2007use}
M.~Gogolla, F.~B{\"u}ttner, and M.~Richters, ``{USE}: a {UML}-based
  specification environment for validating {UML} and {OCL},'' \emph{Science of
  Computer Programming}, vol.~69, no.~1, pp. 27--34, 2007.

\bibitem{Liu2005}
Z.~Liu, H.~Jifeng, and X.~Li, ``{rCOS}: refinement of component and object
  systems,'' in \emph{Proceedings of the 3th Formal Methods for Components and
  Objects (FMCO'04)}.\hskip 1em plus 0.5em minus 0.4em\relax Springer, 2005,
  pp. 183--221.

\bibitem{Jifeng2006}
H.~Jifeng, X.~Li, and Z.~Liu, ``{rCOS}: a refinement calculus of object
  systems,'' \emph{Theoretical Computer Science}, vol. 365, no.~1, pp.
  109--142, 2006.

\bibitem{Ke2012}
W.~Ke, X.~Li, Z.~Liu, and V.~Stolz, ``{rCOS}: a formal model-driven engineering
  method for component-based software,'' \emph{Frontiers of Computer Science},
  vol.~6, no.~1, pp. 17--39, 2012.

\end{thebibliography}

\begin{IEEEbiography}[{\includegraphics[width=1in,height=1.25in,clip,keepaspectratio]{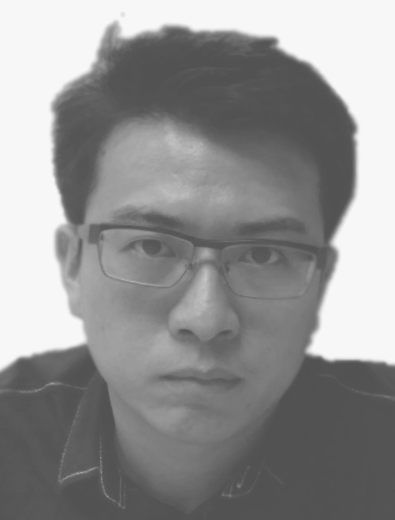}}]{Yilong Yang}
received his B.S. degree in Computer Science from China University of Mining and Technology, China in 2010. His M.S. degree is from Guizhou University, China in 2013. He has been a follow at United Nations University - International Institute for Software Technology, Macau. Currently, he is a Ph.D. candidate in Software Engineering at University of Macau. His research interests include Automated Software Engineering and Machine Learning. 
\end{IEEEbiography}

\begin{IEEEbiography}[{\includegraphics[width=1in,height=1.25in,clip,keepaspectratio]{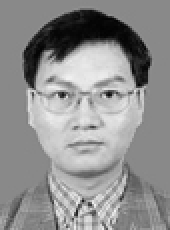}}]{Xiaoshan Li}
received his Ph.D. degree from the Institute of Software, the Chinese Academy of Sciences, Beijing in 1994. Currently, he is an Associate Professor in the Department of Computer and Information Science at University of Macau. His research interests are Health Information Exchange, Formal Methods, Object-oriented Software Engineering, Real-time Specification and Verification, and Semantics of Programming Language.
\end{IEEEbiography}

\begin{IEEEbiography}[{\includegraphics[width=1in,height=1.25in,clip,keepaspectratio]{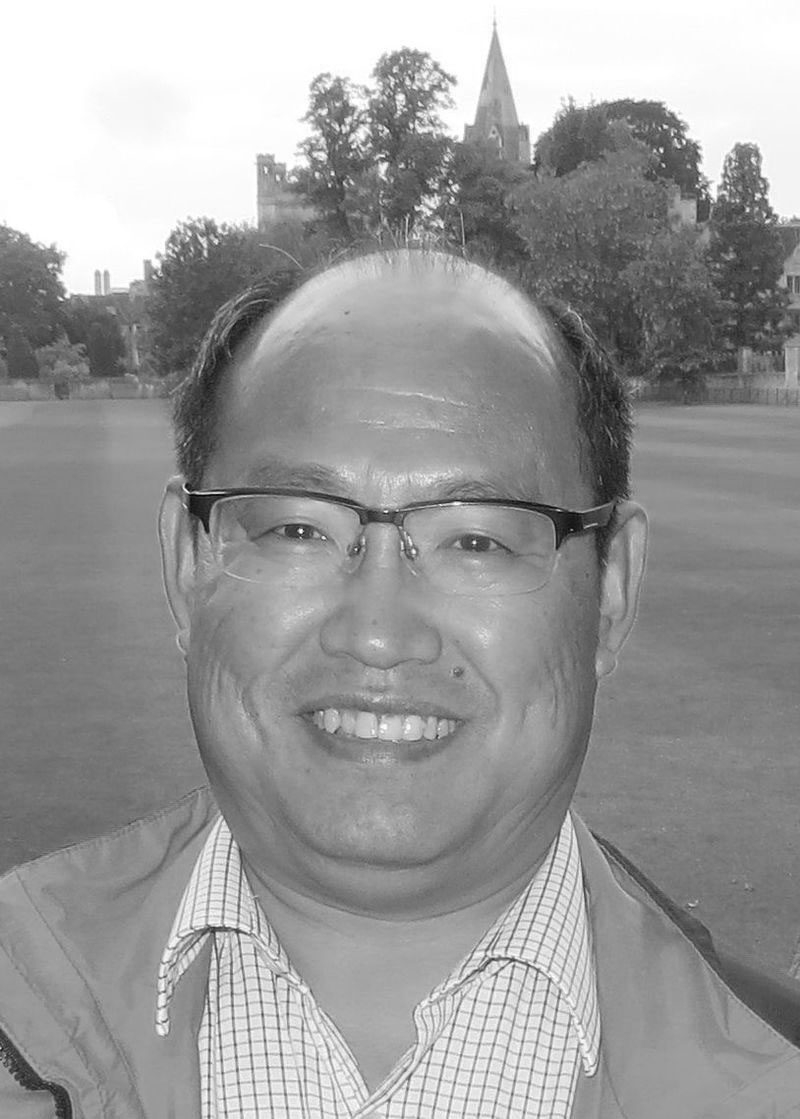}}]{Zhiming Liu}
holds Ph.D. degree in Computer Science from University of Warwick, UK. He worked as a research fellow and then senior researcher at the United National University, International Institute for Software Technology (UNU-IIST, 2002-2013), and Chair Professor of Software Engineering at Birmingham City University (2013-2015). He currently leads RISE, the Centre for Research and Innovation in Software Engineering at Southwest University under the Innovative Talents Recruitment Program (1000-Expect Program). His main research interest is in the areas of Formal Methods, including Real-time Systems, Fault-tolerant Systems, Health Information Systems, and Object-oriented and Component-based Systems.
\end{IEEEbiography}

\begin{IEEEbiography}[{\includegraphics[width=1in,height=1.25in,clip,keepaspectratio]{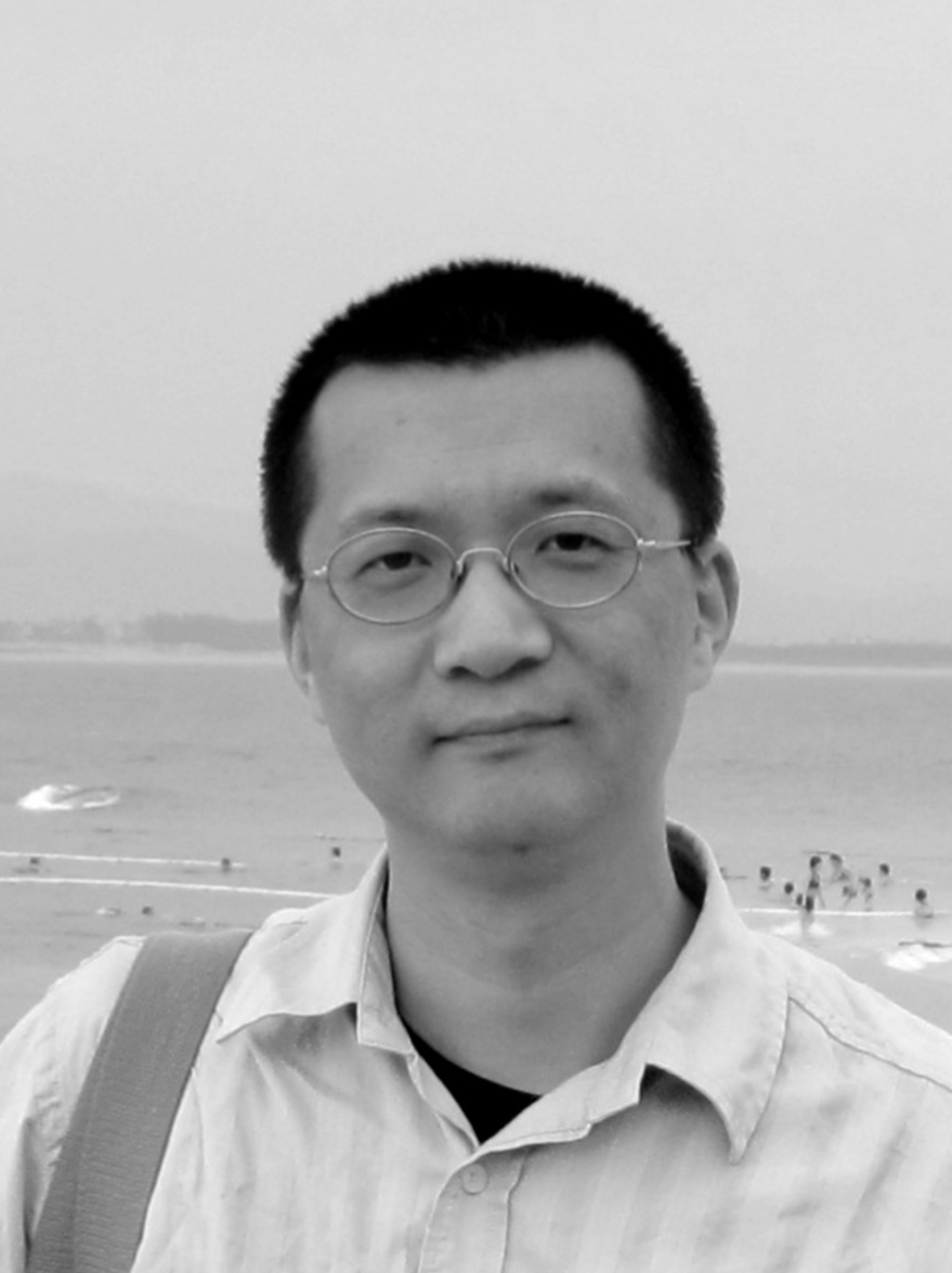}}]{Wei Ke}
received his Ph.D. degree in Computer Applied Technology from Beihang University in 2012. He is currently an associate professor in School of Public Administration of Macau Polytechnic Institute. He had successfully completed in a couple of research projects funded by Macau FDCT. His research interests include Programming Languages, Formal Methods, Software Engineering. 
\end{IEEEbiography}

\begin{IEEEbiography}[{\includegraphics[width=1in,height=1.25in,clip,keepaspectratio]{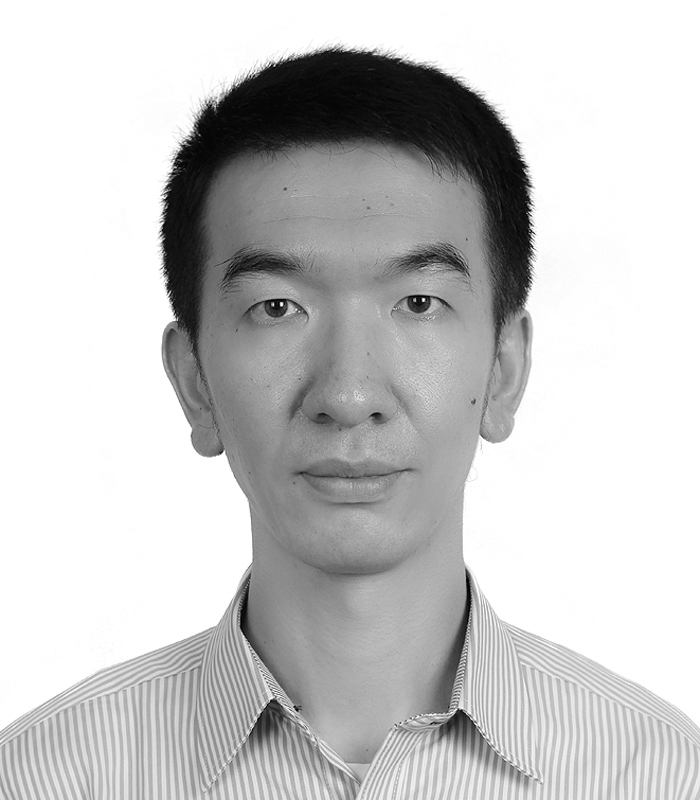}}]{Quan Zu}
got his PhD degree in College of Electronics and Information Engineering, Tongji University, China in 2016. Currently, he is a Post-doctoral Research Fellow in Department of Computer and Information Science of the Faculty of Science and Technology in University of Macau. His research interests include Formal Methods, Model Checking, Algorithm Design and Analysis.
\end{IEEEbiography}

\begin{IEEEbiography}[{\includegraphics[width=1in,height=1.25in,clip,keepaspectratio]{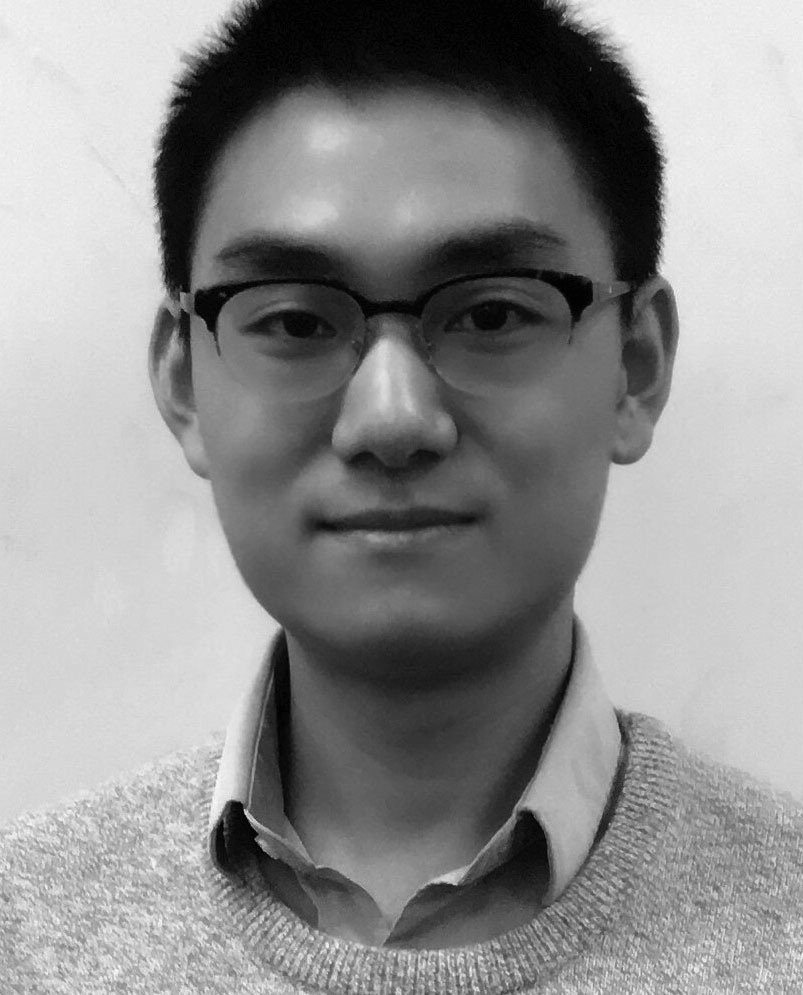}}]{Xiaohong Chen}
is currently a Ph.D. student at University of Illinois at Urbana-Champaign. Before that, he obtained his Bachelor's degree majoring in mathematics at Peking University, China. Xiaohong's main research interest includes Formal Methods and Programming Languages, in particular, Formal Specification and Verification and Program Logics.
\end{IEEEbiography}

\end{document}